\documentclass[preprint,authoryear,11pt,leqno]{elsarticle}

\usepackage[utf8]{inputenc} 		% applemac or utf8
\usepackage[T1]{fontenc}
\usepackage{lmodern}
\usepackage{amssymb,amsmath,stmaryrd}
\usepackage{amsthm}
\usepackage{dsfont}
\usepackage{longtable,booktabs}
\usepackage{enumerate}
\usepackage{multicol}
\usepackage[breaklinks,%
					linktocpage,%
					colorlinks,%
					backref=page]{hyperref}
\hypersetup{colorlinks=true,citecolor=magenta,urlcolor=blue}
\usepackage{epsfig}
\usepackage{pstricks}
\usepackage{graphicx}
\usepackage{algorithm}
\usepackage{algorithmicx,algpseudocode}
\usepackage{fullpage}
\usepackage{lscape}
\usepackage{multirow}
\usepackage{subcaption}

\input epsf

\newcommand{\ba}{\begin{eqnarray}}
\newcommand{\ea}{\end{eqnarray}}

\newenvironment{Proofc}[1]{\smallskip\par\noindent\textbf{#1}\quad}%
  {\hfill$\Box$\bigskip\par}

\nonstopmode

\usepackage[pagewise,displaymath,mathlines]{lineno}

\graphicspath{{figures/}}

\usepackage[mathlines]{lineno}

\journal{Engineering Applications of Artificial Intelligence}

\usepackage{ulem}

\begin{document}
\begin{frontmatter}


\title{Physics Informed Data Driven model for Flood Prediction: Application of Deep Learning in prediction of urban flood development}



\author{Kun Qian, Abduallah Mohamed and Christian Claudel}

\address{Texas, United States}

\begin{abstract}
Flash floods in urban areas occur with increasing frequency. Detecting these floods would greatly help alleviate human and economic losses. However, current flood prediction methods are either too slow or too simplified to capture the flood development in details. Using Deep Neural Networks, this work aims at boosting the computational speed of a physics-based 2-D urban flood prediction method, governed by the Shallow Water Equation (SWE). Convolutional Neural Networks(CNN) and conditional Generative Adversarial Neural Networks(cGANs) are applied to extract the dynamics of flood from the data simulated by a Partial Differential Equation(PDE) solver. The performance of the data-driven model is evaluated in terms of Mean Squared Error(MSE) and Peak Signal to Noise Ratio(PSNR). The deep learning-based, data-driven flood prediction model is shown to be able to provide precise real-time predictions of flood development. Implementation codes can be found in the link\footnote{\url{https://github.com/Kunqian123/DeepFloodPrediction}}.
\end{abstract}

\begin{keyword}
Deep Learning \sep Flood Prediction \sep Shallow Water Equation                                                                               

% MSC codes here, in the form: \MSC code \sep code
% or \MSC[2008] code \sep code (2000 is the default)

\end{keyword}

\end{frontmatter}

% \linenumbers

\begin{abstract}
Floods are attacking urban area more and more frequently. A real-time flood system which can reflect the detailed flood development and provide on-time prediction of flood can help alleviate the loss caused by floods. The present flood prediction methods are either too slow or too simplified to capture the flood development in 2-D sense. This work boosts the computation speed of a physics-based flood prediction method which is governed by Shallow Water Equation(SWE) and is able to predict the flood development in 2-D sense with Deep Neural networks. Convolutional Neural Networks(CNN) and Generative Adversarial Neural Networks(GANs) are applied to learn the dynamics from the data simulated by a Partial Differential Equation(PDE) solver. 
\end{abstract}
\section{Introduction}
\label{sec:intro}

% \kun{Add a figure here to explicitly show the Purpose of the article. The figure: on the left, the inputs pictures, in the middle, three kinds of methods/or only the ensemble method, on the right, the output values.}

% In the introduction part, the following things need to be mentioned:
% Background:
% \begin{itemize}
%     \item the more and more frequent flood happening around the world request more techniques to monitor the flood to reduce the harm of flood.
%     \item the PDE based method is too low in solving to meet up the requirement of real time monitoring.
%     \item the development in deep learning techniques show the possibility to learn a flood dynamics which is governed by a very nonlinear PDE equation.
% \end{itemize}
% \begin{itemize}
%     \item we are going to use deep learning method to learn the dynamics of flood to predict the flood status every $5$ minutes
%     \item the data of flood studied will be based on the shallow water equation instead of history data.
%     \item we are going to learn the dynamics of flood in 2-D sense, and we will not focus on the discrete cases. That's to say, we will learn do a pixel-to-pixel learning.
%     \item Austin area will serve as the background of the example.
% \end{itemize}

Floods are the most common disaster worldwide \cite{berz2000flood}. Every year, floods cause more than $6,000$ casualties, and have an estimated cost of $25$ billion dollars \cite{berz2000flood}. Several flood mitigation strategies exist, including flood channels, early warning systems \cite{krzhizhanovskaya2011flood}, temporary flood barriers, etc. \cite{few2003flooding}. Among all these strategies, flood monitoring is one of the most inexpensive, and most resilient strategy to uncertainties \cite{montz2002flash}. Hence, timely, accurate early warning systems could lead to a dramatic reduction of flood-caused injuries and fatalities. Given the very large areas that require monitoring, mobile sensing platforms, such as Unmanned Air Vehicles (UAVs), are most suitable for such problems. UAVs have been used for a variety of monitoring applications, including air pollution monitoring in~\cite{alvear2017using} and flood monitoring in~\cite{mohammed2014uavs, dunbabin2012robots}.

\begin{figure}[!ht]
\begin{center}
\includegraphics[width=\textwidth]{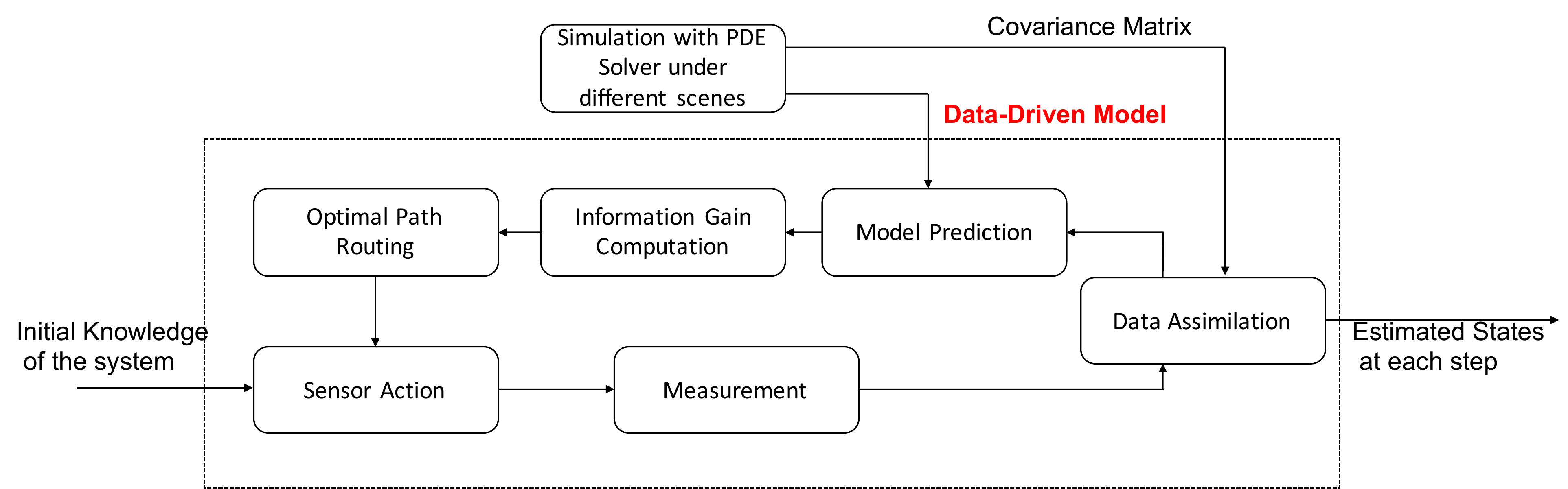} 
\caption{Overview of the estimation process for a real-time flood monitoring system. The focus of the present article is the data-driven model.}
\label{fig:estimationprocess}
\end{center}
\end{figure}

Accurate, real-time prediction of flash flood onset and progression is a challenging problem. It requires the fusion of measurements with computationally-intensive flood propagation models to estimate the water levels and velocities over an entire region. Modern estimation techniques are mainly based on a Bayesian Tracking procedure with two steps: Prior Update and Measurement Update. These are conducted recursively to update the estimated state of the system \cite{bergman1999recursive}. This process is summarized in  Fig.\ref{fig:estimationprocess}. The first step, prior update, depends on the prior knowledge that people have about the system and its dynamics. The update can be executed using either (1) a physics-based model describing the propagation of water flows or (2) using purely statistical methods. Statistical methods usually require historical data. Lots of researches in this field apply time-series prediction models to collected data from fixed gauge stations to predict flood or river stream flows \cite{khac2012data, damle2007flood, xiong2001non, laio2003comparison, valipour2012parameters, valipour2013comparison, basha2008model}. Neural Networks and its variants are also frequently used to extract the stream fluctuation rules from historical data and predict future floods \cite{coulibaly2000daily, sattari2012performance, elsafi2014artificial, humphrey2016hybrid, sit2019decentralized}. Interested readers can refer to \cite{mosavi2018flood} for a detailed review for the application of machine learning techniques to flood predictions. However, statistical prediction methods can only give a broad prediction for river systems or flood basins. Furthermore, the lack of historical data in urban areas makes it troublesome to use statistical methods to address the issue of real-time urban flood prediction. Instead, physical models which depend on Partial Differential Equations (PDEs) can simulate, thus predict, flood development given the realistic settings.

Flood dynamics are usually governed by the Shallow Water Equation (SWE) \cite{vreugdenhil2013numerical}. With its origin in the Navier-Stokes Equation, SWE is widely applied to environmental flow simulation. The general characteristic of shallow water flows is that the vertical dimension is much smaller than the horizontal scale. Although SWE shows a 3-D attribute due to the bottom friction \cite{vreugdenhil2013numerical}, in most cases, it's sufficient to average over the vertical horizon to get the 2-D SWE. Since \cite{alcrudo1993high}, the finite volume method is popular for the numerical simulation of SWE. The main progresses of the numerical solutions for SWE are listed as follows. \cite{anastasiou1997solution} extended the finite volume based numerical solution of SWE to unstructured mesh. \cite{kurganov2002central, audusse2004fast, kurganov2007second, liang2009numerical} contributes to the 'well-balanced' numerical simulation of SWE which can preserve numerical stability of the dry area and still water. People also worked out the well-balanced solution of the wet-dry front problem in SWE ~\cite{horvath2015two, bollermann2013well}.

Recent research probes implementing SWE numerical simulation on Graphics Processing Units (GPU) \cite{brodtkorb2012efficient}. Although PDE-based models can give detailed prediction for flood events in urban areas, the computation speed of the SWE solver is too slow to meet the real-time prediction requirements, especially for city-wide simulations. It can be even worse if the Monte-Carlo methods, such as the Ensemble Kalman Filter(EnKF) which is widely applied in large scale environmental systems, are used in estimation. Present physics-based real-time prediction methods rely heavily on simplifications. The 2-D domain is simplified to the catchments and links between them in which only 1-D Saint-Venant equations need to be solved ~\cite{yakowitz1985markov, georgakakos1986generalized, moore1994rffs, bellos2016hybrid, bout2018validity}. The simplifications lead to imprecise predictions and cannot cover every corner of a city-wide area.

Recently, deep learning has risen as a promising way to extract highly non-linear relations from data \cite{lecun2015deep}. With its structured architecture and automatic training process with back propagation \cite{rumelhart1988learning}, researchers are using deep learning to solve PDEs \cite{raissi2018deep, raissi2017physics1, raissi2017physics2, sirignano2018dgm} or boost the speed of PDE solvers \cite{Tompson2017AcceleratingEF, tompson2017accelerating,thuerey2018well, Wiewel2018LatentspacePT, xie2018tempogan, chu2017data}. Through processing the data simulated by PDE solvers, deep learning models can mimic the non-linearity carried by PDEs with a much faster computation speed in prediction. \cite{thuerey2018well, Wiewel2018LatentspacePT, xie2018tempogan, chu2017data} follow this idea and apply different deep learning methods to problems related to the Navier-Stokes equation. However, these studies only focus on theoretical settings and haven't been extended to large scale PDE systems.

% This work focuses on a test area around Austin, Texas, illustrated in Fig. \ref{fig:awsmall} and \ref{fig:studyarea}. 

% The difficulty arises as the PDE solver, especially for large scale spatial domains, is computationally intensive. As an example, the forward simulation of a prior, over a time horizon of $10$ minutes, cannot be finished within $10$ minutes with a regular laptop computer. Since ensemble-based estimation methods used to estimate flood propagation require the forward simulation of a large number of ensembles during the prediction step, these physics-based models can only be used for real-time estimation with massive amounts of computational power. In this article, our primary objective is to quickly and accurately compute the future state of the system, through leveraging the physics of flood development into a quick to compute Deep Neural Network. The general procedure is shown as \ref{fig:overvoew}. 

\begin{figure}
    \centering
    \begin{subfigure}[b]{0.4\textwidth}
        \centering
        \includegraphics[width = \textwidth]{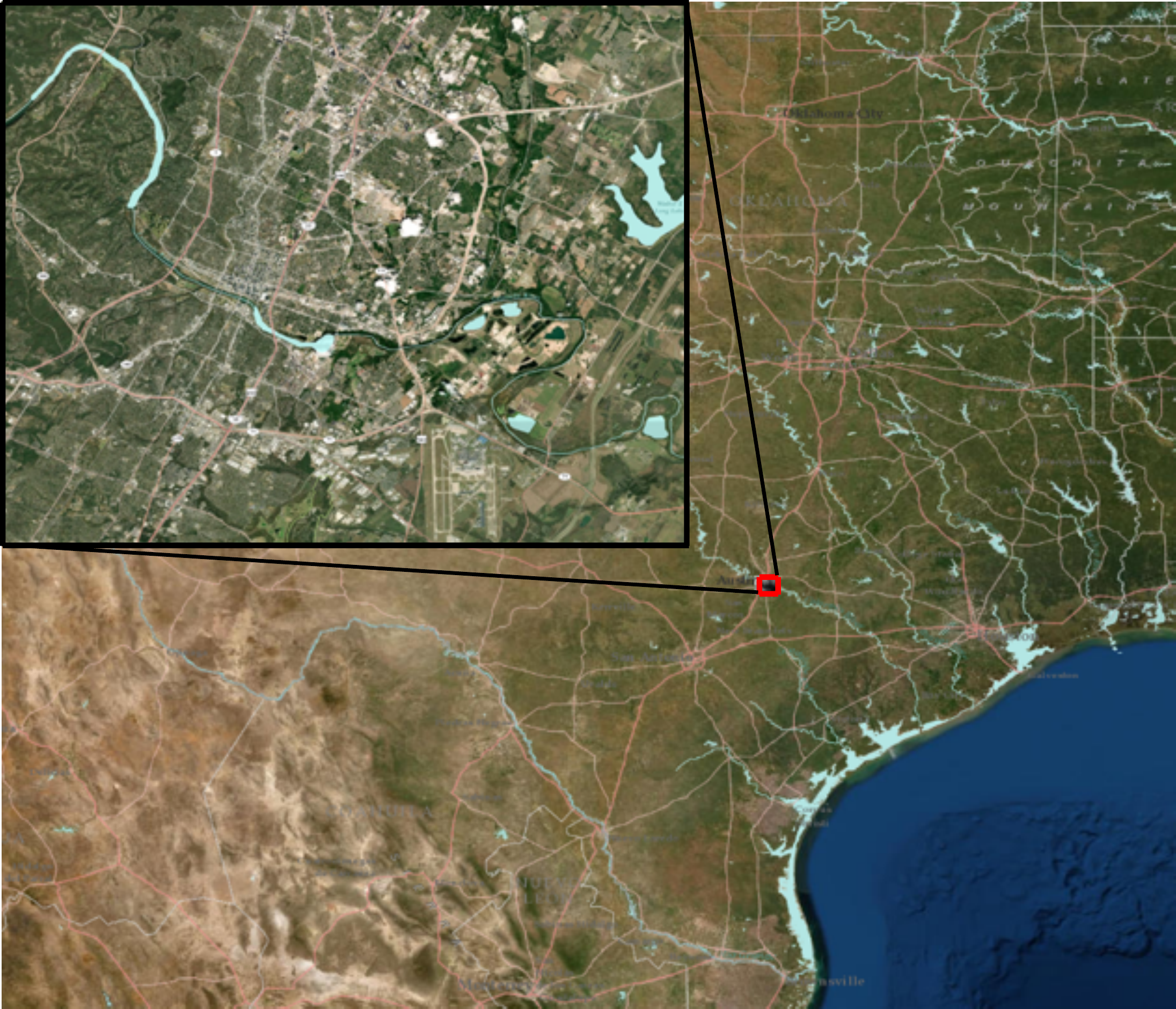}
        \caption{The location of Austin.}
        \label{fig:studyarea}
    \end{subfigure}
    \hfill
    \begin{subfigure}[b]{0.5\textwidth}
        \centering
        \includegraphics[width = \textwidth]{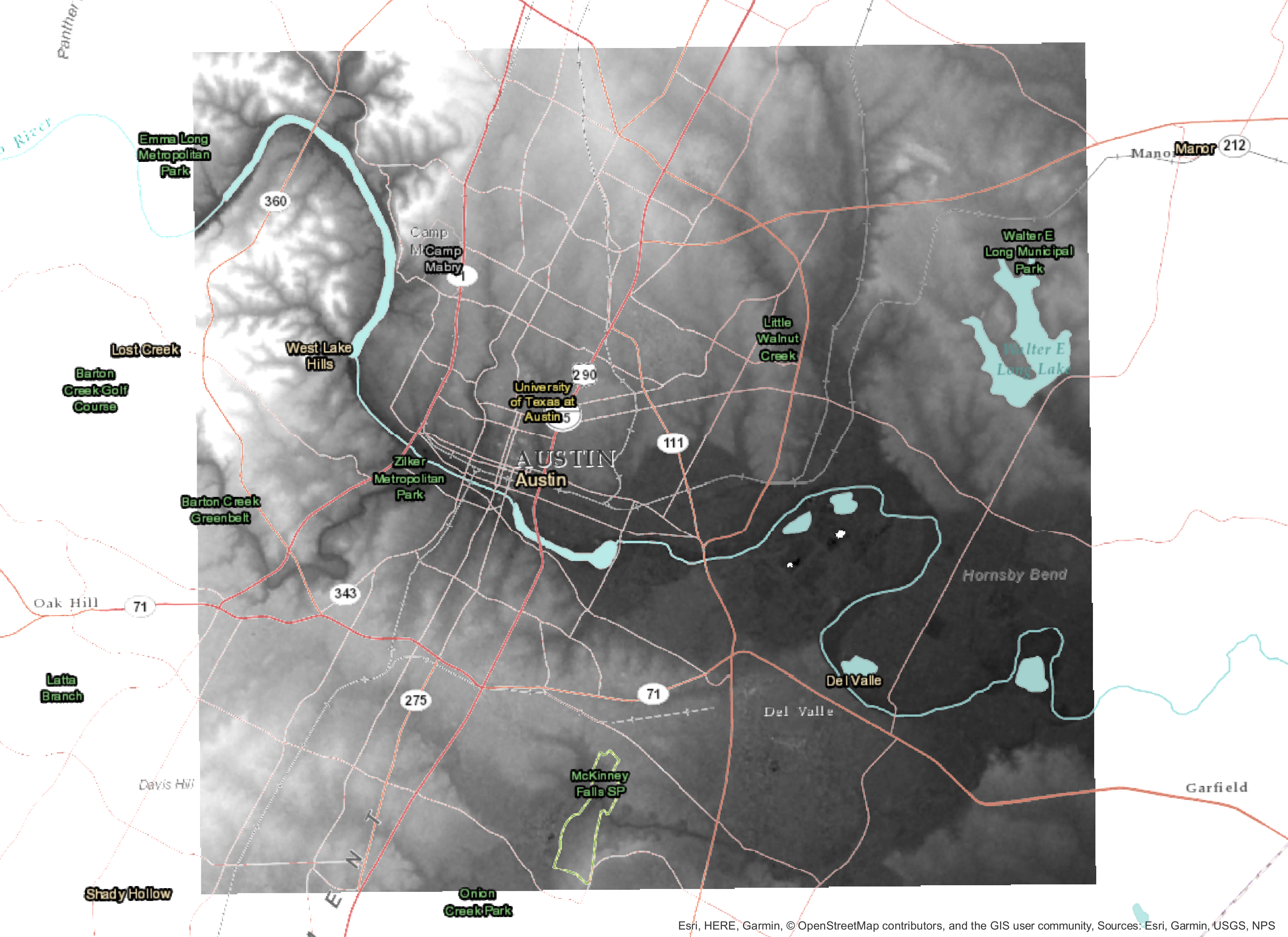}
        \caption{The topology of Austin.}
        \label{fig:awsmall}
    \end{subfigure}
    \caption{Map of the studied area. The left subfigure represents the subset of the state of Texas that we study in this article. The right subfigure represents an elevation map of this area.}
\end{figure}

The objective of this study is to investigate the application of deep learning to boost SWE solver speeds for the purpose of real-time flood prediction. Specifically, given an initial state and boundary conditions, the data-driven model will predict the future states subject to various inputs. Austin, Texas is targeted as the area receiving the flood (shown in Fig.\ref{fig:studyarea} and Fig.\ref{fig:awsmall}). The deep learning based, data-driven model will carry the physics governed by SWE equations and the parameters that imitate the features of the Austin area to make real-time predictions on floods. The precise data-driven model can replace SWE models to provide real-time and detailed prediction of flood evolution on a city-wide scale. To the best of our knowledge, this is the first real-time flood prediction model which can give detailed prediction of a two dimensional flood over a large spatial domain. This is also the first study that investigates deep learning to reconstruct the dynamics of the 2D SWE in a realistic urban setting.

% We\kun{Can I use 'we' here?} expect that the dynamics governed by SWE, the influence of boundary conditions, the parameters of the PDE (including the topology of the studied area, the friction coefficients of land, gravitational constant, the physical parameters of water) can be completely captured by the large amount of simulation data. From the regression of the data, the Deep Learning model can inherit the dynamics governed by SWE and the influence of the boundary conditions and parameters. 

The organization of this paper is as follows. Section \ref{sec:methodology} talks about the methodology of the study. Section \ref{sec:dataprepare} introduces the procedure to produce data for deep learning. The performance of deep learning based, data-driven models are evaluated and discussed in Section \ref{sec:result}. Section \ref{sec:conclusion} summarizes the study and the obtained results. Potential future works are discussed. More results and training details of the neural networks are presented in the Appendix.

% The applications of the learned deep learning models can be applied to prediction with real-time requirement. In this work, we apply our work to the real-time monitoring of flood in real-time within the Area of Austin, TX \ref{fig:awsmall}, though we will only focus on how to get the desired deep learning model and will not cover the estimation part. We show in the last section of this article that the resulting deep-learning model is robust to the measurement update thus is able to be applied to estimation. 
% \input{tex/2_Relatedwork.tex}
\section{Methodology}
\label{sec:methodology}
% \begin{itemize}
%     \item Firstly, give a general idea of the form of PDE to show that it can be expressed by nonlinear regression methods such as deep learning.
%     \item Then give the detailed expression of shallow water equation in 2-D sense to show our input and output of deep learning.
%     \item argue that this is a markov process so that we only need $t$ step to predict step $t+1$
%     \item talk about the kind of deep learning structure we are going to employ: CNN.
% \end{itemize}
% In this chapter, general methodology is focused on and the Deep Learning techniques applied are introduced. 

\subsection{General Procedure}

\begin{figure}[!ht]
\begin{center}
\includegraphics[width=\textwidth]{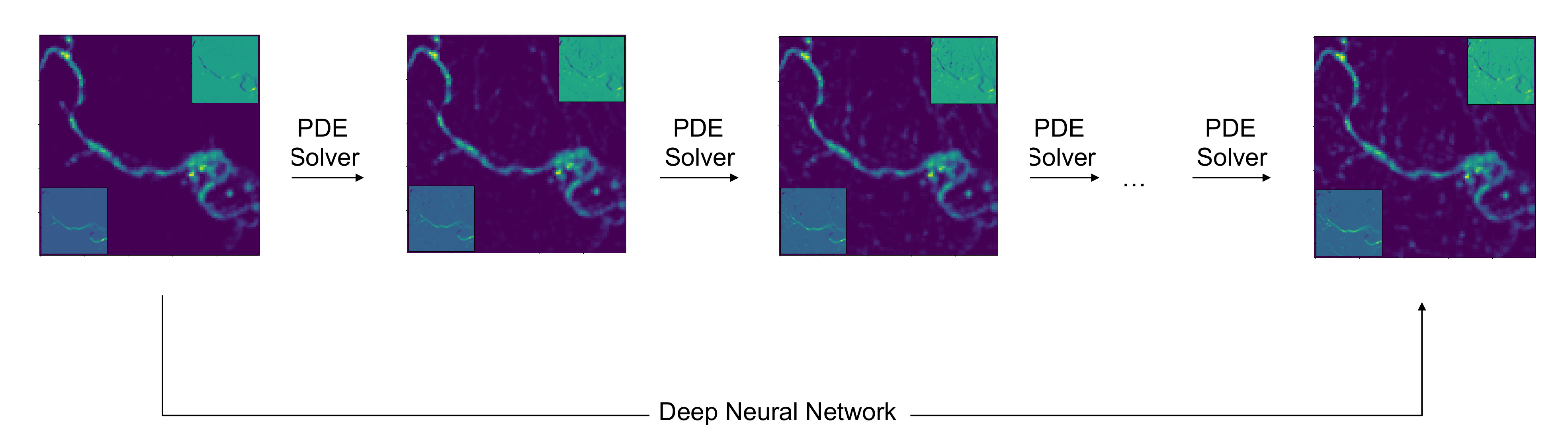} 
\caption{Overview of the objective of this work. With PDE solver, thousands of computation steps are needed to predict over a horizon of $5$ minutes. In this work, the objective is to replace the PDE solver with a deep learning model which can directly predict the states several minutes later with one step of computation to realize the real-time prediction task.}
\label{fig:overview}
\end{center}
\end{figure}

This work proposes a deep neural network that can substitute the SWE-based solver for predicting the evolution of floods. The philosophy behind this idea is simple. The deep neural network is trained on large quantities of data simulated by the SWE simulator with different initial states, inputs and boundary conditions. In numerical simulations, time steps are set to be small to achieve high precision and numerical stability while also dragging down the computation speed. The data-driven model is set to predict over a larger time step to boost the prediction speed. The general process is illustrated in Fig.\ref{fig:overview}. Difficulties arise from the high dimensional inputs and outputs in learning, highly non-linear dynamics by SWE and the gaps in data scale between cases.

% The general assumption of the SWE is that the water depth is much smaller comparing to the other spatial scales of problem. In this case the fluid speed in depth is averaged to get rid of the vertical dimension. Interested readers are refered to\cite{vreugdenhil2013numerical} for more details about the derivation of SWE from the Navier Stokes Equation. 
Starting from SWE, this section will go through the construction process of the data-driven model including inputs, outputs and deep learning techniques applied.

\eqref{equ:SWE} is the incompressible shallow water equation in 2D conservative form, obtained from~\cite{tan1992shallow, zoppou1999catastrophic}. 

\begin{equation}
\label{equ:SWE}
\frac{\partial \mathbf{U}}{\partial t} + \bigtriangledown \mathbf{F} = \mathbf{S}
\end{equation}

where $\mathbf{U}$ is the vector of conservative variables, $\mathbf{F}$ is the flux tensor and $\mathbf{S}$ is the source term. Changing Equ. \eqref{equ:SWE} to Cartesian form will result in Equ. \eqref{equ:SWE-cartesian}.

\begin{equation}
\label{equ:SWE-cartesian}
\frac{\partial \mathbf{U}}{\partial t} + \frac{\partial \mathbf{E}}{\partial x} + \frac{\partial \mathbf{G}}{\partial y} = \mathbf{S}
\end{equation}

where $\mathbf{E}$ and $\mathbf{G}$ are cartesian components of $\mathbf{F}$. The detailed expression of $\mathbf{U}$, $\mathbf{E}$, $\mathbf{G}$ and $\mathbf{S}$ are in \eqref{equ:details}.

\begin{equation}
\mathbf{U} = 
\begin{bmatrix}
h\\
uh\\
vh
\end{bmatrix} \text{, } \mathbf{E} = 
\begin{bmatrix}
uh\\
u^2h + gh^2/2\\
uvh
\end{bmatrix}\text{, }\mathbf{G} = 
\begin{bmatrix}
vh\\
uvh\\
v^2h + gh^2/2\\
\end{bmatrix}
\label{equ:details}
\end{equation}

% \begin{equation}
% \mathbf{E} = 
% \begin{bmatrix}
% uh\\
% u^2h + gh^2/2\\
% uvh
% \end{bmatrix}
% \end{equation}

% \begin{equation}
% \mathbf{G} = 
% \begin{bmatrix}
% vh\\
% uvh\\
% v^2h + gh^2/2\\
% \end{bmatrix}
% \end{equation}

Different force terms can be included in the term $\mathbf{S}$ depending on the chosen application. Here, gravity and the friction between water and ground are included as two basic terms.

\begin{equation}
\mathbf{S} = 
\begin{bmatrix}
0\\
gh(S_{g_{x}} - S_{f_{x}})\\
gh(S_{g_{y}} - S_{f_{y}})
\end{bmatrix}
\end{equation}
in which $S_{g}$ only depends on the gravity coefficient and the slope of the topology. With $\eta$ as friction coefficient, $S_f$ is defined according to Manning resistance law \cite{gioia2001scaling} as $S_{f_x} = \frac{u\eta \sqrt{u^2 + v^2}}{h^{4/3}} $ and $S_{f_y} = \frac{v\eta \sqrt{u^2 + v^2}}{h^{4/3}} $.

According to the above equations, $u$, $v$ and $h$ are the three state variables of a system governed by SWE. Let us denote $u$, $v$ and $h$ at time step $t$ as $u^t$, $v^t$ and $h^t$. Let us denote the $u$, $v$ and $h$ values of discrete domain as vector $\mathbf{\bar{U}}$, $\mathbf{\bar{V}}$ and $\mathbf{\bar{H}}$. Suppose $\mathbf{\phi}$ is a numerical method derived from Equ.\ref{equ:SWE-cartesian} such that
\begin{equation}
\label{equ:onestep}
\begin{bmatrix}
\mathbf{\bar{U}}^{t+\delta t}\\
\mathbf{\bar{V}}^{t+\delta t}\\
\mathbf{\bar{H}}^{t+\delta t}
\end{bmatrix} = 
\mathbf{\phi}(\begin{bmatrix}
\mathbf{\bar{U}}^{t}\\
\mathbf{\bar{V}}^{t}\\
\mathbf{\bar{H}}^{t}
\end{bmatrix}, p^t)
\end{equation}

in which $p^t$ is the input to the system at time step $t$. Various numerical methods can be used to solve Equ. \ref{equ:SWE-cartesian}, though each corresponding $\phi$ will be either computationally expensive or the associated time step $\delta t$ will have to be small. As a result, the total number of computations required to solve Equ. \ref{equ:SWE-cartesian} is huge and its execution usually takes a long time, especially in cases of large scale watersheds.

The general idea of this work is to find a deep learning model $\Phi$ that can provide precise and fast prediction of the future states of a system governed by the SWE, given the present states and forecast inputs. The function $\Phi$ is related to $\phi$ according to the following expression:
% For example, when doing a real time monitoring of flood using UAVs, we only need to have a prediction of the future states over a longer time step, for example five minutes or ten minutes ahead in the future is suitable to provide guidance for our mobile monitoring system. Both explicit and implicit PDE models suffers from slow inference which is not suitable for real-time application. By using deep models with a small amount of parameters it is possible to overcome this issue. 
\begin{equation}
\begin{split}
\label{equ:deep_regression}
\begin{bmatrix} 
\mathbf{\bar{U}}^{t+n\delta t}\\
\mathbf{\bar{V}}^{t+n\delta t}\\
\mathbf{\bar{H}}^{t+n\delta t}
\end{bmatrix} &= 
\mathbf{\Phi} (
\begin{bmatrix}
\mathbf{\bar{U}}^{t}\\
\mathbf{\bar{V}}^{t}\\
\mathbf{\bar{H}}^{t}
\end{bmatrix},p^{t},\dots,p^{t+(n-1)\delta t})\\
& = \mathbf{\phi}(\mathbf{\phi}(\dots \phi(\begin{bmatrix}
\mathbf{\bar{U}}^{t}\\
\mathbf{\bar{V}}^{t}\\
\mathbf{\bar{H}}^{t}
\end{bmatrix}, p^t),\dots, p^{t+(n-2)\delta t}), 
p^{t+(n-1)\delta t})
\end{split}
\end{equation}

It is noticeable that only the present states $\begin{bmatrix}
\mathbf{\bar{U}}^{t}\\
\mathbf{\bar{V}}^{t}\\
\mathbf{\bar{H}}^{t}
\end{bmatrix}$ is included in right side of Equ.~\ref{equ:onestep} and Equ.~\ref{equ:deep_regression}. As the SWE is a first order PDE and can be written in homogeneous form \cite{zoppou1999catastrophic}, it has the Markov Property which means only present states are needed to predict future states. Thus, it is not necessary to depend on previous information. Although Recurrent Neural Network (which keeps historical information) can still be applied, Convolutional Neural Network will be enough for this task. Moreover, it was shown in \cite{bai2018empirical} that Convolution operations can also be applied to temporal prediction tasks.

% In Equ.\ref{equ:deep_regression}, the terms $\begin{bmatrix}
% \mathbf{U}^{t-k\delta t}\\
% \mathbf{V}^{t-k\delta t}\\
% \mathbf{H}^{t-k\delta t}
% \end{bmatrix}$ are for the numerical approximation of the term $\frac{\partial \mathbf{U}}{\partial t}$ in Equ.\ref{equ:SWE-cartesian}. However, as the $n$ increases which means that $\mathbf{\Phi}$ will predict over a large time step, the 

% The computation speed of function $\Phi$ is fast. Also, the computational speed can be further boosted with setting $n$ a large number as long as the size of prediction gap satisfies real-time monitoring requirement.

% Since this method is data-driven,the data comes from numerous numerical simulation results from PDE solvers. $\Phi$ will be drawn from the simulation data by fitting a deep neural network.

It should be noted that only the expression of the dynamics is provided in Equ.~\ref{equ:SWE-cartesian}. In order to properly pose this PDE problem, boundary condition constraints and initial conditions are needed. These constraints will be specified in Section~\ref{sec:dataprepare}.

% Because of the fame in being able to do regression for those extremely nonlinear processes, deep learning can serve as the function $]Phi$. Here are the advantages of using deep learning to serve as function $\Phi$. Firstly, with the arbitrarily defined neural network depth and structure, deep learning can do the regression for any nonlinear process in theory. Secondly, the deep learning model allow us to do the vectorized computation in prediction. Because when we are applying particle filter[citation needed] and ensemble kalman filter[citation needed], we need to usually compute hundreds of different states. The vecorized computation can help save large amount of time.

% Thus the simple forward structure such as Convolutional Neural Network(CNN) could be able to capture this relation and there is no necessity to apply Recurrent Neural Network(RNN) based methods. In fact, even for a system without Markov Property which means that the prediction depend on several previous steps, it has been argued that temporal CNN can perform as well as Recurrent Neural network such as LSTM and GRU \cite{bai2018empirical}. Different from previous work in Deep Learning and PDEs, this work shows that given the Markov Property of the problem, CNN can give a satisfying regression result for PDEs.

\subsection{Convolutional Neural Network}
Since the data representation to the network is two-dimensional, which contains $3$ channels each pixel, Convolutional Neural Networks (CNN)s \cite{lecun1998gradient} are suitable for such a data representation. Convolutional layers, which apply the same parameters over each local sub-part of the images, mimic the convolution operation. This type of model greatly reduces the number of parameters needed in the neural network.
% There have been lots of attempts in using deep learning to do nonlinear regression for Partial Differential Equations[acclerated  eulerian...][physics informed deep learning...], many of them apply a pyramid shape structure.(a picture needed to show this structure). However, their settings of partial differential equations, such as boundary conditions, are quite theoretical and simple. As we are going to study the Austin area with the real topology of Austin area, the dynamics will be different in each local area as the height, and slope differs. Thus, if we just follow a common convolutional way in which we use the same kernels of parameters to apply to every elements, it's not that reasonable given the condition that we don't know about the topology information when learning.

\begin{figure}
    \centering
    \begin{subfigure}[b]{0.45\textwidth}
        \centering
        \includegraphics[width = \textwidth]{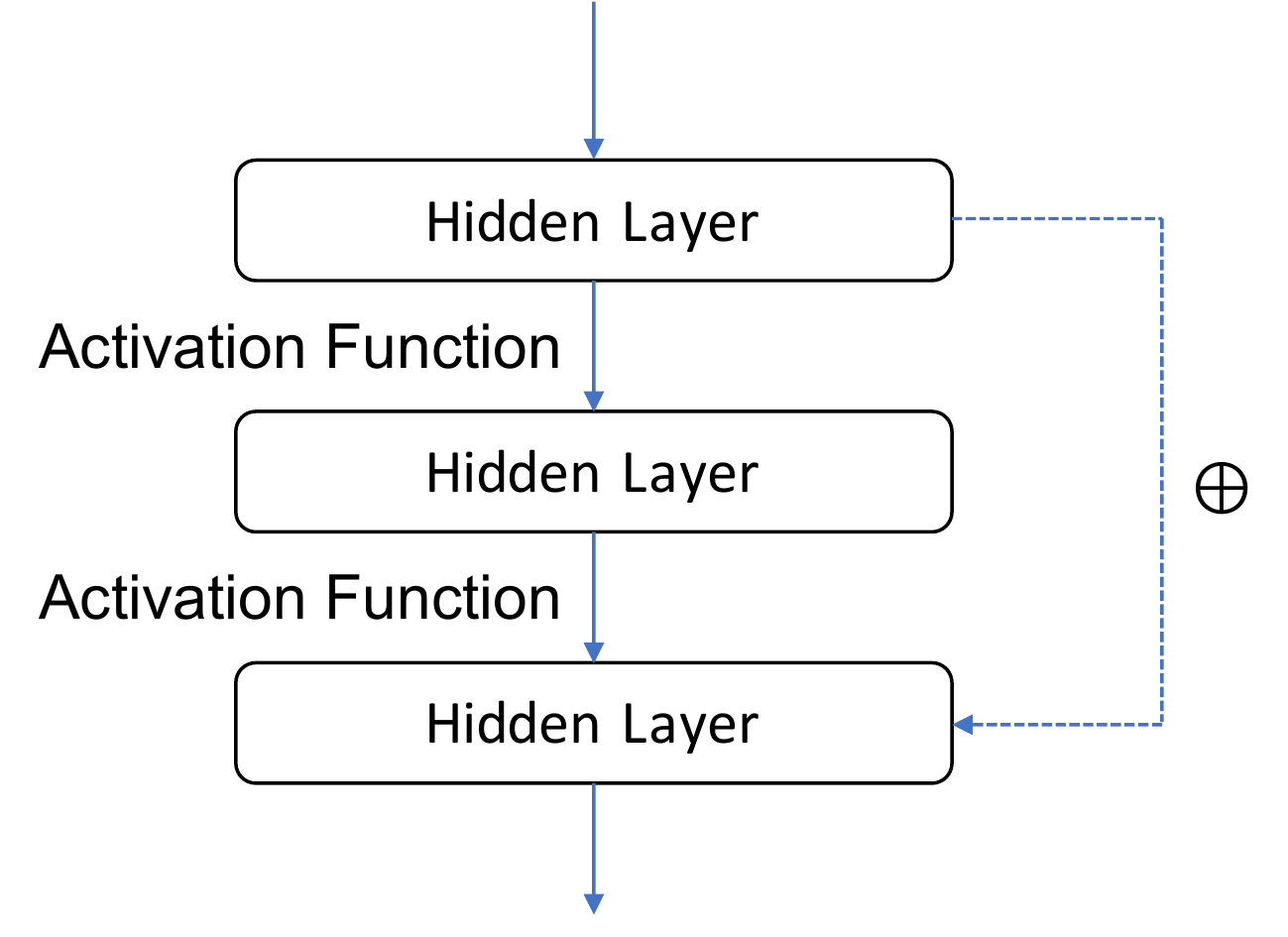}
        \caption{Residual Block}
        \label{fig:residual}
    \end{subfigure}
    \hfill
    \begin{subfigure}[b]{0.45\textwidth}
        \centering
        \includegraphics[width = \textwidth]{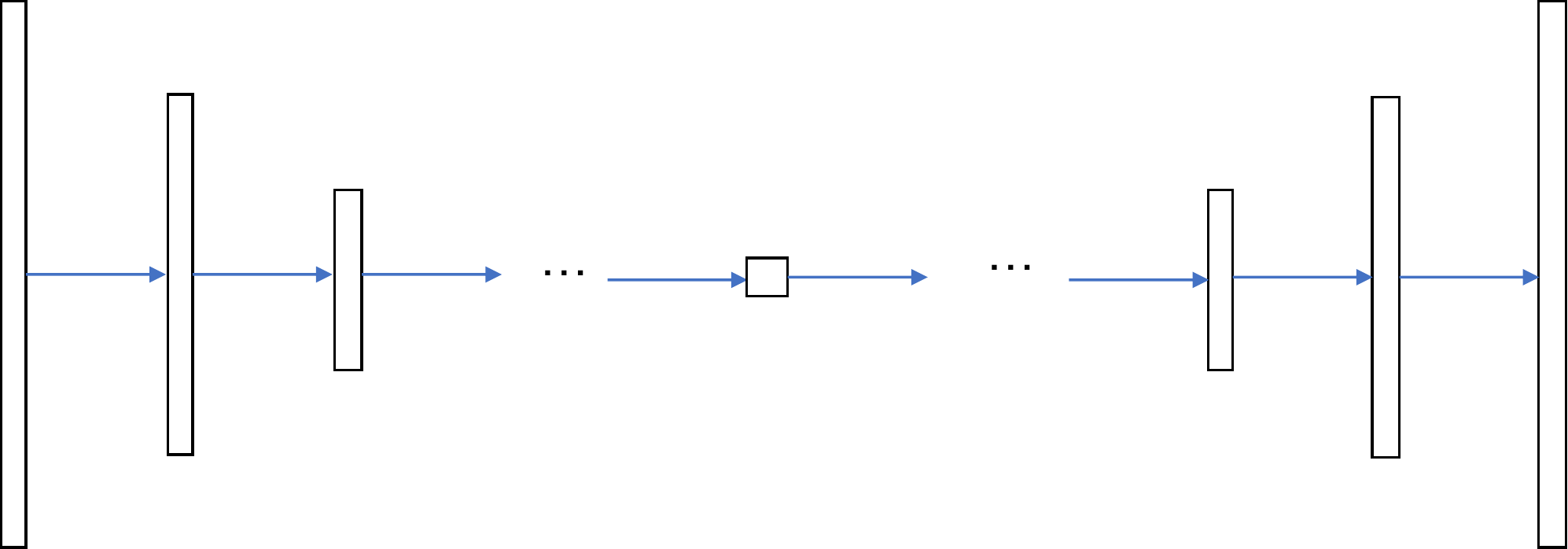}
        \caption{Network Architecture illustration}
        \label{fig:UNet}
    \end{subfigure}
    % \caption{}
    \label{fig:nnarchitecture_demo}
\end{figure}

For this pixel-to-pixel regression task, a neural network architecture is applied such that the dimension of the input in each channel is first reduced while the number of channels increases. The dimension of each channel is then increased while the number of channels is reduced. The idea is illustrated in Fig.~\ref{fig:UNet}. In order to avoid vanishing gradient in deep networks, the idea of residual block extracted from ResNet~\cite{he2016deep} is applied among the network. Residual blocks (Fig.\ref{fig:residual}) have been shown to ease the learning process of deep neural networks in general, owing to their ability to overcome the problem of decaying gradients. The $L_1$ norm loss is used in the loss function. $L_1$ norm loss can measure the absolute difference between predictions and targets. Comparing to $L_1$ norm loss, $L_2$ norm tends to give blurred prediction in pixel-to-pixel tasks \cite{mirza2014conditional} and thus is not employed.

% Another choice is to first reduce the dimension in the feature layers and then enlarge the dimension back to the output dimension again. This process can share similar structures to tasks such as encoders and decoders. In this work, we mainly exploring these two kinds of architectures and their variations. We tested over ten different structures for each kinds. The general architecture is shown below Fig \ref{fig:UNet}. Detailed information will be introduced in the later sections.

\subsection{Generative Adversarial Networks}

Generative Adversarial Networks (GANs), introduced in~\cite{goodfellow2014generative}, is a class of deep models that learn the distribution of the data. Usually, GANs consist of two deep models, a generator ${G(z;\theta_g)}$ that attempts to generate a realistic sample from the data distribution $\sim p_{\text{data}}$ using random noise ${p_z(z)}$ as an input, and a discriminator ${D(x; \theta_d)}$ that classifies these samples  ${p_g}$ as real or fake. 

\begin{equation}
\label{equ:minimaxgame-definition}
\min_G \max_D V(D, G) = \mathbb{E}_{{x} \sim p_{\text{data}}({x})}[\log D({x})] + \mathbb{E}_{{z} \sim p_z({z})}[\log (1 - D(G({z})))].
\end{equation}\

Similarly to CNNs, Generative Adversarial networks (GANs) can also conduct pixel-to-pixel regression with conditionsl GANs. As we attempt to predict the next state of the flood based on the current state, conditional GANs \cite{mirza2014conditional} are particularly suited. In conditional GANs, we condition the input and the output with a sample from our initial states of the flood ${{f}}$. The GANs value function becomes: 

\begin{equation}
\label{eq:minimaxgame-definition-conditioned}
\min_G \max_D V(D, G) = \mathbb{E}_{{x} \sim p_{\text{data}}({x})}[\log D({x} | {f})] + \mathbb{E}_{{z} \sim p_z({z})}[\log (1 - D(G({z} | {f})))].
\end{equation}

\begin{figure}
    \centering
    \includegraphics[width=\textwidth]{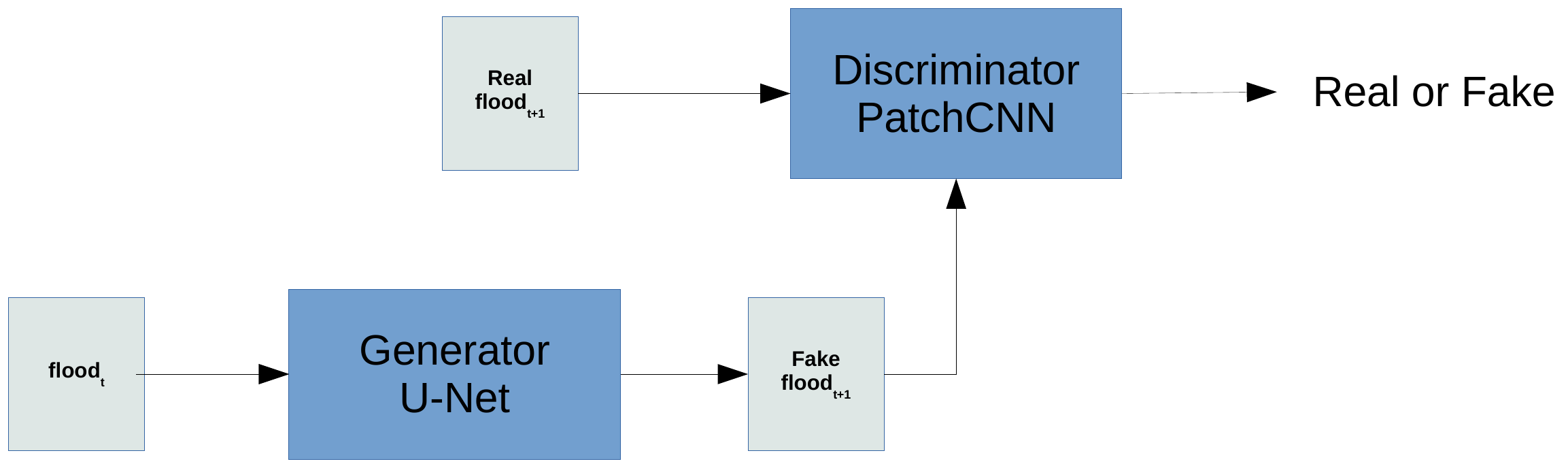}
    \caption{The GAN workflow used for flood prediction.}
    \label{fig:GAN_process}
\end{figure}

The process of the conditional GANs is illustrated in Fig.\ref{fig:GAN_process}.\\

% \kun{change the figure a little bit}

In fact, conditional GANs can be viewed as improving objective functions beyond CNNs, from a fixed $L1$ norm loss to a trainable neural network which can capture the structural error~\cite{dosovitskiy2016generating, li2016combining, mirza2014conditional}. In order to pursue both a small structural loss and absolute error between predictions and targets, $L_1$ norm loss and neural network based discriminator are usually mixed together in loss functions. Notice that the conditional GANs will go back to CNNs when the relative weight of $L_1$ norm loss is set high.

\subsection{Improving the performance of conditional GANs with Kalman Filter}
% \kun{This section is super mathematically unsound, but I am strugling to find a proof for it. Here is the original motivation for this. What I am thinking is to how to ensemble two prediction method. This must be considered from a probability view. }

% As we have introduced, the motivation for the data-driven model here is to meet the real time estimation requirement. The data driven model would be applied to the estimation task. Consider the case that the measurement is usually point-wised, and estimation techniques such as Kalman Filter is applied, we would like to introduce a bit about Kalman Filter and an ensemble method to improve the prediction precision of these methods based on the form of Kalman Filter. In later numerical experiment section, we would like to validate if the data-driven model can be applied to estimation task. 

GANs are suitable to generating predictions following a certain type of distribution. However, this may lead to poor performance in flood prediction. At the situation of light rains (small scale inputs), the changes in flood states are so subtle that the distribution can be assumed to be static. Thus, sometimes conditional GANs cannot predict subtle changes at the beginning stage. This will cause a divergence between predictions and targets. Two ways are proposed to solve this problem. First, the weight of $L_1$ norm loss in loss functions can be increased to care more about subtle changes. Second, an assimilation method is raised to overcome the possible divergence of conditional GANs based models, the procedure of which is similar to the measurement update step of Kalman Filter. The details of the second method is illustrated below.

Introduced by \cite{kalman1960new}, Kalman Filter is a classical estimation technique based on two steps. The second step, referred to as the Measurement Update, can be a method of merging information from two different sources: an a-priori estimate and measurement data. Depending on the co-variance matrix, the Kalman Filter can update the system with partial observations: 

% \begin{equation}
% \label{equ:kalmangain}
%     K = \mathbf{P}\mathbf{H}^T (\mathbf{H}\mathbf{P}\mathbf{H}+\mathbf{R})^{-1}
% \end{equation}
\begin{equation}
\label{equ:measurementupdate}
    \mathbf{\mathbf{\hat{X}}} = \mathbf{X} + \mathbf{P}\mathbf{H}^T (\mathbf{H}\mathbf{P}\mathbf{H}+\mathbf{R})^{-1}(\mathbf{z} - \mathbf{H}\mathbf{X})
\end{equation}

\begin{figure}
    \centering
    \includegraphics[width = \textwidth]{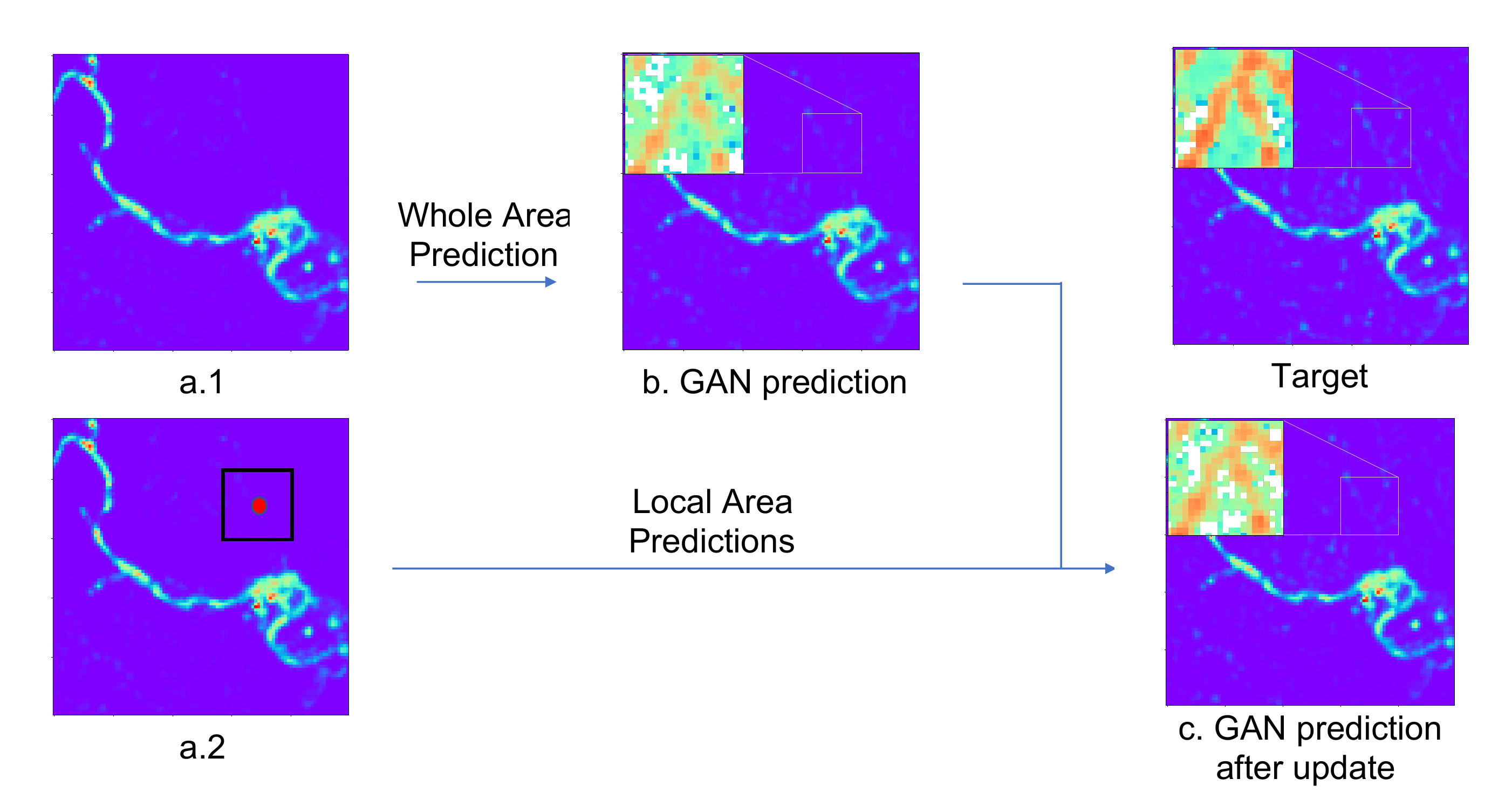}
    \caption{The demonstration of how we depend on Kalman Filter to improve the prediction performance of cGANs. a.1 and a.2 are the same input. We use a.1 and cGANs to give a whole-area prediciton(b). We use local areas(black box) in a.2 to predict the states at red dot places. Then, the prediction at red dot places are used to improve the prediction precision of the local area of b.GAN prediction. The improved result is shown in c. When predicting, we depend on around 50 spots similar to the red spot in a.2.}
    \label{fig:ensemblecompare}
\end{figure}

The assimilation method takes exactly the same form as Equ.\ref{equ:measurementupdate}, though the meaning of the variables differs. Given the prediction $\mathbf{X}$ by conditional GANs and the prediction $\mathbf{z}$ by some other prediction models, $\mathbf{P}$ and $\mathbf{R}$ are the co-variance matrices of prediction error of $\mathbf{X}$ and $\mathbf{z}$ respectively. Suppose $z$ is allowed to be a partial prediction, the matrix $\mathbf{H}$ connects the prediction $\mathbf{X}$ and $\mathbf{z}$. The performance of this kind of ensemble method is shown in Fig.\ref{fig:ensemblecompare}. In practice, $\mathbf{z}$ can be predicted by some small neural networks which only predicts states with a small number of specific pixels. The predictions by small neural networks are expected to be more precise than conditional GANs in those pixels. With the co-variance matrix $\mathbf{P}$, surrounding areas are also updated by this step.

\section{Data Preparation}
\label{sec:dataprepare}

In this section, the process of generating training data is detailed. To focus on learning the SWE dynamics with deep learning, the training data is generated with the following assumptions:
\begin{itemize}
    \item The rain rate (input of the model) is set before the simulation (\emph{i.e.~not forecast by this data-driven model}).
    \item The soil properties are constant over the simulation period (hence the model is not taking into account the change of the runoff coefficient as the soil absorbs water).
    \item The influence of the sewers in this area is not included in the model.
\end{itemize}
 
It is stated in Section \ref{sec:methodology} that water depth and water momentum in the x, y directions can represent a system governed by SWE. Combining this with the above assumptions, the three states can be extended to represent flood states in urban areas.

% The main focus of this article is the ability for deep neural networks to approximate SWE under different initial states, precipitation distribution patterns and boundary flood levels. Since we do not consider the soil infiltration and sewer conditions in this work, we can totally capture the state of flood with surface water depth, and surface water momentum. For more precise modeling, additional states (not included as such in the SWE) should be included, including soil state, underground water level, sewer water level should definitely be included to fully represent the system status.

\begin{figure}
     \centering
     \begin{subfigure}[b]{0.5\textwidth}
         \centering
         \includegraphics[width=\textwidth]{figures/aw_small_1.pdf}
         \caption{Elevation map of Austin, TX.}
         \label{fig:awsmalltopo}
     \end{subfigure}
     \hfill
     \begin{subfigure}[b]{0.4\textwidth}
         \centering
         \includegraphics[width=\textwidth]{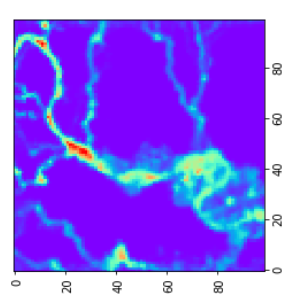}
         \caption{Water level map associated with a simulated flood example.}
         \label{fig:simulationdemo}
     \end{subfigure}
    %  \caption{}
     \label{fig:simudemo}
     \caption{This figure illustrates an example flood simulation. Left: illustration of the topology (the Colorado river is superimposed to the elevation map for clarity). Right: example flood simulation result using the SWE with a set rain rate function and inflows from river channels.}
\end{figure}

% \subsection{Data Generation}
% \begin{itemize}
%     \item GIS data of Austin area(here, show a picture)
%     \item the simulation platform that I used(ANUGA) and the simulation method ANUGA uses(finite volume method). The area size, number of volumes divided, forces considered, water input should be briefly introduced.
% \end{itemize}

\begin{figure}
    \centering
    \includegraphics[width=\textwidth]{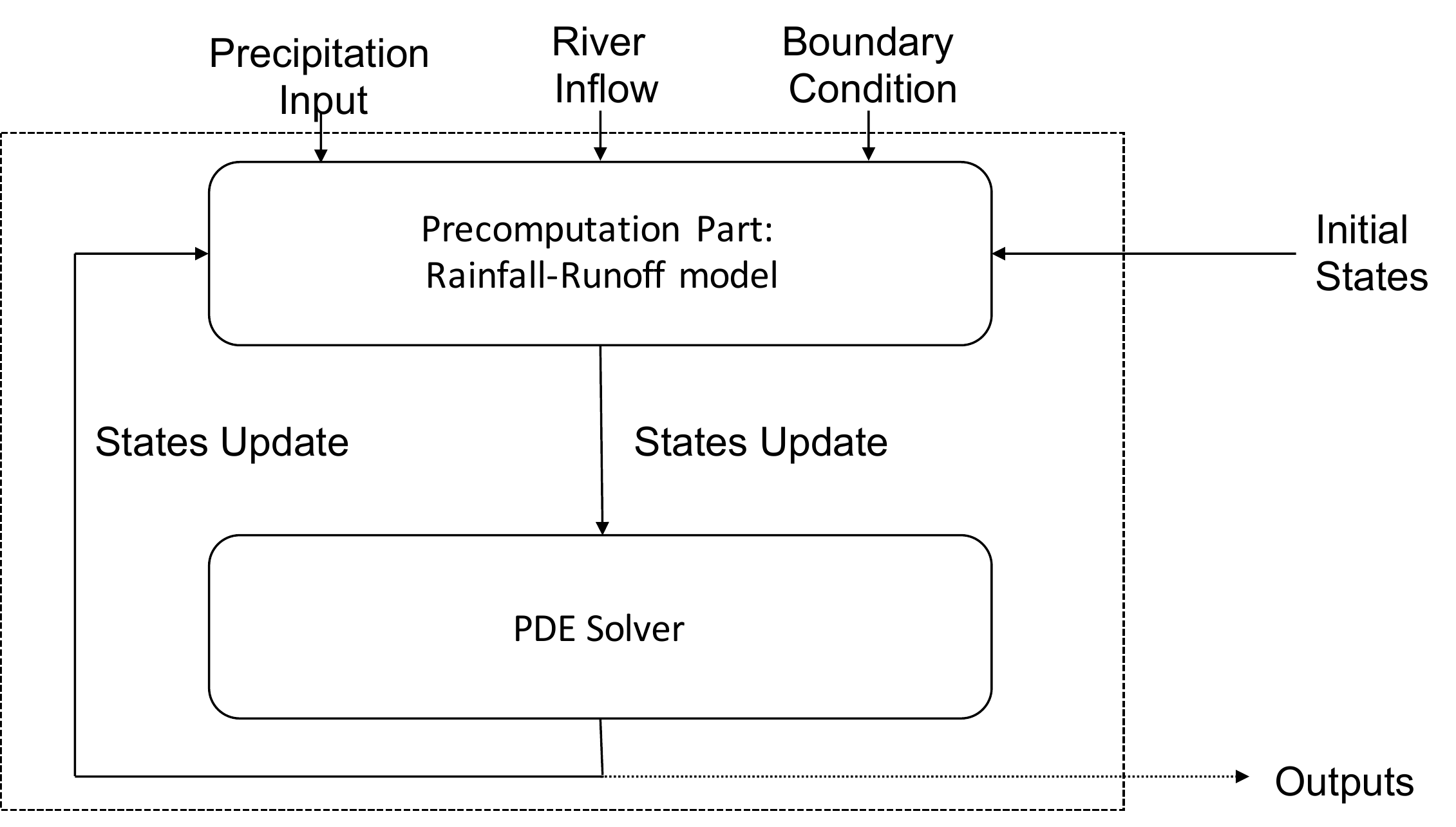}
    \caption{Computational process used for forward simulation of floods using the SWE.}
    \label{fig:simulationprocess}
\end{figure}

The numerical simulations are done in Python and consist of two steps. The first step is to compute the input map over time (rain rate), and the boundary condition functions, used to feed the PDE solver. The second step is to run the PDE solver with a set initial state, the computed boundary conditions, and the input functions. The process is shown in Fig.~\ref{fig:simulationprocess}.

% In this work, we have made some assumptions on the numerical settings in the first part of computation. In the first part, we simply distribute the precipitation evenly across the area. In a more realistic model, the soil infiltration model should be considered, thus more states should be added to the system states. Also, evaporation is not considered in this work. However, as a preliminary work, we focus on the core of the problem-approximating the Shallow Water Equation. 

% The core PDE solver used to solve the shallow water equation requires  the current level of surface water depth and water momentum, in addition to the rain forcing. The expected output is the updated surface flood states.

The open-sourced SWE solver ANUGA Hydro developed by~\cite{ANUGA} serves as the SWE solver in the simulation. It employs a finite volume scheme to solve the 2-D SWE. Inside ANUGA, the mesh generator Triangle from \cite{shewchuk1996triangle} is employed to mesh the area into Delaunay triangulations. The areas of interest (basin and creek areas) are finer meshed than other areas that are not likely to flood. Using the finite volume method, ANUGA applies the numerical scheme from \cite{kurganov2001semidiscrete} to approximate the numerical flux function. The time step size of the simulation is chosen to satisfy Courant–Friedrichs–Lewy (CFL) condition \cite{courant1928partiellen} for stability.

As mentioned in Section \ref{sec:intro}, Austin, Texas is set as the test location. The topology of this area is shown in Fig.\ref{fig:awsmall}. The studied area is divided into $9$ sub-areas, and it is assumed that the precipitation density is spatially uniform within a sub-area. Besides precipitation, inflows from river channels are another input to the model. For the boundary conditions of the studied domain, static water level conditions are assumed, representing Dirichlet boundary conditions. Between each simulation, the level of boundary water depth is different and chosen randomly within a reasonable range. Although the boundary flood level is not included in the inputs of the deep learning models, it is assumed that the flood depth near the boundary area can reflect the boundary water level. Friction coefficients are chosen according to the type of soil in Austin area, using tabulated values.

% The mesh used in the PDE solver has a direct influence on the quality of the simulation results. In our simulations, irregular meshes are applied and the entire area is meshed into irregular triangles. It is worth noticing that the mesh is finer in the river and channel area comparing to other areas, as the gradients are important in river and channel areas.

A total of $1080$ different simulation configurations (combinations of different precipitation patterns, precipitation amount and inflow patterns) are simulated. In each case, the flood is simulated over an horizon of $12$ hours. For convenience, the data generated on the irregular triangle mesh is resampled onto a regular grid using interpolation. The data is thus projected on a $100 \times 100$ grid, saved every $5$ minutes of simulation time.

When extracting data from the simulation results to train the neural network, a stored result is randomly taken as an input and after $30$ minutes, the simulated state serves as the output. The precipitation and inflows between these two time steps are extracted and averaged, as are the other two input channels. In total $10,000$ pairs of inputs and outputs are extracted for training, and another $2,000$ as validation data.

To summarize, each input is an image with $5$ channels of dimension $100 \times 100$. The five channels corresponds to the water depth, water momentum in the $x$ direction, water momentum in the $y$ direction, the expected inflow in the next $30$ minutes, and the expected precipitation in the next $30$ minutes. The output data is a $100\times 100$ image with $3$ channels, representing the state of system $30$ minutes after the input. The training data can be found in the link \footnote{\url{https://drive.google.com/open?id=1F_yts1zp1srwVS1LsGFM-Rch4npbRB_Z}}.

% We prepare the above result as a $5 \times 100 \times 100$ input data, and the target data is formed as $3 \times 100 \times 100$. The $5$ input channels are flood depth, flood momentum in X direction, flood momentum in Y direction, water amount added by precipitation, water amount added because of river inflow. The $3$ output channels are flood depth, flood momentum in X direction and flood momentum in Y direction. Normalization is conducted in each channel. We normalize the data to between the range $[-1, 1]$ for the ease of training.
% (a picture of the area that we studied is needed here)
% Details about the simulation settings and the explanation of data can be found in this link.(Write another file to explain data and the settings in the simulation).

% \subsection{Data Normalization}

\section{Results and discussion}
\label{sec:result}

\begin{figure}
    \centering
    \includegraphics[width = \textwidth]{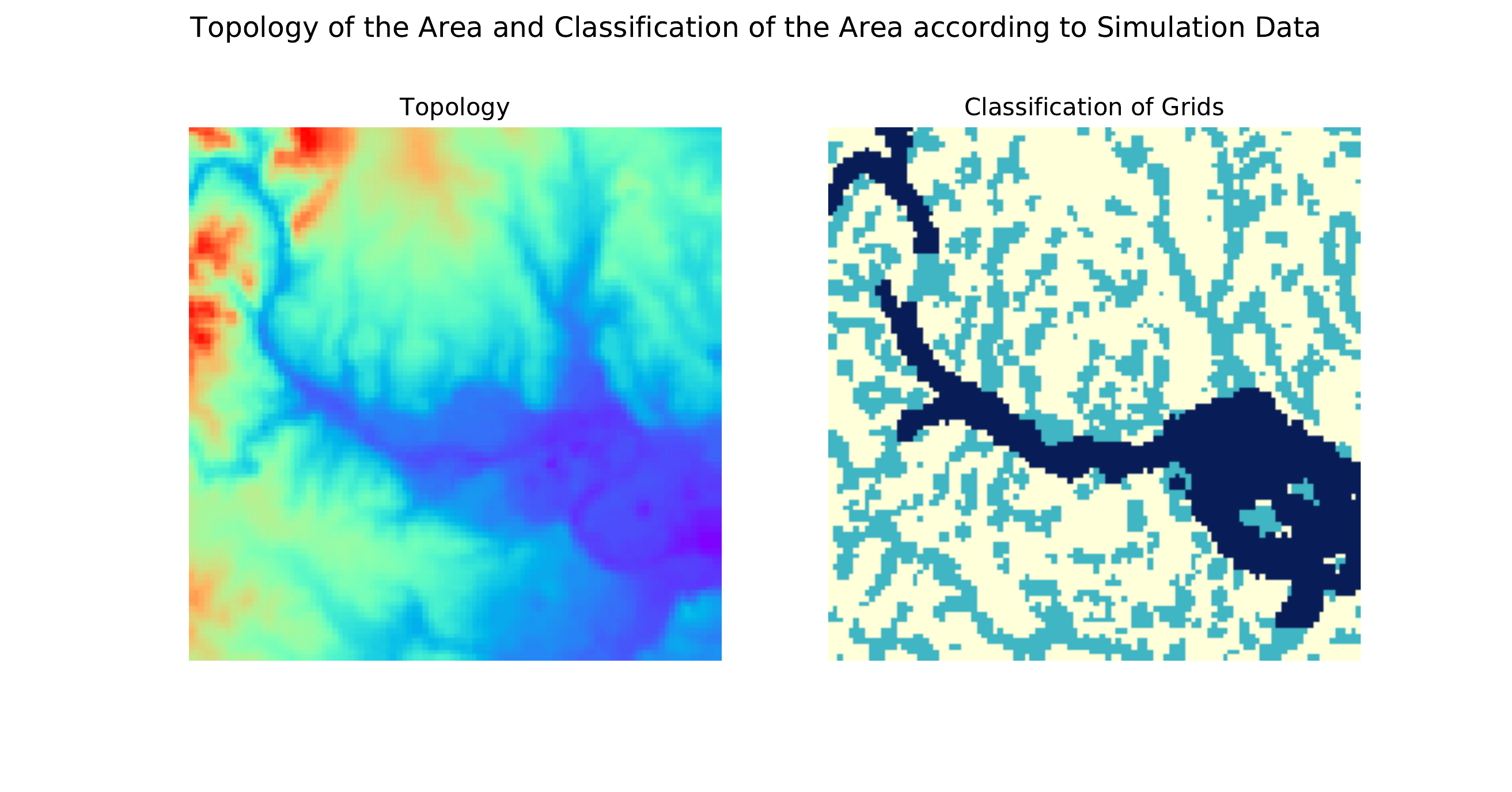}
    \caption{The Classification of the area.}
    \label{fig:topo_class}
    \caption{The left subfigure represents the elevation of each grid point in our study area. The right subfigure indicates the different classes: dark grid points are the Colorado river basin, light blue areas are small channels, and white areas are dry lands.}
\end{figure}

In order to better analyze the results of the regression, pixels in the studied domain are divided into three classes according to simulation data using K-Means~\cite{macqueen1967some}. The three classes respectively correspond to the Colorado River (the main river), small creeks and channels (which can flood or not), and permanently dry areas. In the Colorado river area, the water depth tends to be large and the speed of water is relatively stable. In creeks and channels, there are considerable variations in water depth and velocity during flood development. This is the area that people are most interested in. In other areas, grids are always dry, irrespective of the simulated scenario. The three classes are illustrated in Fig.~\ref{fig:topo_class}.

The rest of this section serves to compare models trained under different settings. With the same CNN architecture (the optimal one), Model 1 adds noise to the inputs during the training process, while Model 2 does not. Model 3 is a conditional GANs based model, in which the discriminator only has a relatively small weight comparing to $L_1$ norm component in loss function. Model 3 also adds noise in the training inputs. Model 4 and Model 5 are conditional GANs based models in which the weight of the discriminator is higher than the $L_1$ norm component in loss function. Models 4 and 5 are identical except for the application of the assimilation step introduced in Section~\ref{sec:methodology} in Model 5.

\subsection{Comparison of different models}

The precision and computation speed in predicting one step forward are evaluated in this subsection. The precision is measured by MSE error and the computation speed is compared based on how much the deep learning models boost the SWE solver.

\begin{table}
\centering
\begin{tabular}{|l|l|l|l|} 
\hline
\multirow{2}{*}{Methods} & \multicolumn{3}{l|}{ Class attributes}    \\ 
\cline{2-4} & River & Channel & Land \\ 
\hline
Model 1   &     0.000376930       &    0.000375208       &        0.000026189          \\ 
% \hline
% CNN with UNet &        0.032496481     &     0.013583293        &      0.001343256         \\ 
\hline
Model 2   &     0.000352408   &    0.000346435     &      0.000020496        \\
\hline
Model 3   &      0.000478783      &    0.0005027733         &      0.00003498       \\
\hline
Model 4   &       0.006085431      &      0.003261132       &       0.000161186       \\
\hline
Model 5   &       0.004006399      &      0.002234181       &       0.000161186       \\
\hline
\end{tabular}
\caption{Precision of one-step prediction (water depth, x-momentum and y-momentum) with different methods in MSE sense. The results are averaged over $2000$ test cases.}
\label{table:precision_general}
\end{table}

\begin{table}
\centering
\begin{tabular}{|l|l|l|l|} 
\hline
\multirow{2}{*}{Methods} & \multicolumn{3}{l|}{ Class attributes}    \\ 
\cline{2-4} & River & Channel & Land \\ 
\hline
Model 1   &     0.000313255       &    0.000150732       &        0.000013969          \\ 
% \hline
% CNN with UNet &        0.061071739     &     0.006991209        &      0.001343256         \\ 
\hline
Model 2  &     0.000367964   &    0.000129624     &      0.00000875        \\
\hline
Model 3   &    0.000417673       &     0.00022312      &    0.0000166921        \\
\hline
Model 4   &       0.003452165      &      0.001244793        &       0.000101648       \\
\hline
Model 5   &      0.002452165      &      0.000837835       &       0.000101648      \\
\hline
\end{tabular}
\caption{Precision of one-step precision (water depth only) with different models, in the MSE sense. The results are averaged over $2000$ test cases.}
\label{table:precision_depth}
\end{table}

As shown in Table \ref{table:precision_general} and Table \ref{table:precision_depth}, CNN-based models generally have higher precision in prediction than those of GANs. Although conditional GANs has shown its advantages over figure translation tasks \cite{mirza2014conditional}, it did not show many advantages over plain CNNs in the pixel-to-pixel regression task in this work.  Meanwhile, it is shown that increasing the weights attached to $L_1$ norm in loss function can improve the performance of conditional GANs. As discussed in Section \ref{sec:methodology}, discriminator in GANs can be viewed as a way to measure structural loss while $L_1$ norm directly measures the absolute difference between predictions and targets. Thus, it makes sense that putting more weights on $L_1$ norm in loss function will reduce the error measured by MSE. Model 5 also shows an improvement over Model 4, which means the assimilation approach can improve the precision of Model 4 albeit not as much as Model 3. Comparing the performance of Model 1 and Model 2, it can be seen that training without added noise will bring us a more accurate result in predicting for only one step. However, the higher precision in predicting one step ahead does not always mean a better performance in predicting the temporal evolution for a long time horizon. This will be discussed further in the next subsection.

\begin{table}
\centering
\begin{tabular}{|l|l|} 
\hline
Methods   &       Computation Speed            \\ 
\hline
PDE Solver  &       $1 \times$           \\ 
\hline
Model 1   &        $50000 \times$           \\ 
\hline
Model 2   &             $50000 \times$            \\
\hline
Model 3  &           $50000 \times$          \\
\hline
Model 4   &    $50000 \times$         \\
\hline
Model 5  &           $10 \times$          \\
\hline
\end{tabular}
\caption{Computational time improvements. All computations are performed on a Intel i7-7700 CPU $@$3.60GHz. Since the computational time of the PDE solver depends on the initial conditions and the inputs of the problem, the PDE solver speed is averaged over a large number of different cases.}
\label{table:speed}
\end{table}

Since the main motivation of developing a data-driven model to substitute the PDE solver is to realize the real-time prediction, the computation speed is compared in table \ref{table:speed} with PDE solver as baseline. To be fair, we conduct the computation of PDE solver and data-driven model on the same computer without parallel computing activated. It can be observed from the Table \ref{table:speed} that data-driven models boost the computation speed greatly and are fast enough to serve as prediction models for real-time flood monitoring. Furthermore, the computational speed of PDE solver is unsteady. It varies considerably with water speed and can be extremely slow when modeling fast water streams. However, the computation speed of the data-driven model is quite stable. This is a significant advantage of data-driven models for estimation tasks. Ensemble Kalman Filter(EnKF), which depends on a large number of ensembles to capture the non-linearity of the system, is usually used in estimating large scale systems. With the improvements in computation speed, the data-driven model makes it possible to apply EnKF to flood estimation tasks.

\subsection{Prediction over temporal evolution}

The performances of data-driven models in predicting temporal evolution are discussed in this section. In the study of numerical simulator for PDEs, stability is a main focus. Extending it to data-driven models, it is concerned whether the data-driven model can stably and realistically reproduce the dynamics carried by the SWE. For all the test cases in this subsection, data-driven models are only shown with the initial conditions and the inputs at each step. The predictions by data-driven models are compared with the predictions by SWE solver.

In order to measure the performance of each method quantitatively, MSE error\footnote{$MSE = \sum_i \sum_j (\hat{x}_{i,j} - x_{i.j})^2$} and PSNR(Peak Signal to Noise Ratio)\footnote{$PSNR = 10\log_{10}(\frac{\sigma_{peak}^e}{MSE})$, $\sigma_{peak}^e$ is the peak signal value.} are computed for each method.

% \kun{show some example pictures of sequences}

% \begin{figure}
%     \centering
%     \includegraphics[width =1 \textwidth]{figures/time_series20_target.pdf}
%     \caption{The PDE simulated temporal Development of a Flood Process}
%     \label{fig:temporal_target}
% \end{figure}

% \begin{figure}
%     \centering
%     \includegraphics[width =1 \textwidth]{figures/time_series20_Res.pdf}
%     \caption{The CNN(Resblock) predicted temporal Development of a Flood Process}
%     \label{fig:temporal_res}
% \end{figure}

% \begin{figure}
%     \centering
%     \includegraphics[width =1 \textwidth]{figures/time_series20_UNet.pdf}
%     \caption{The CNN(U-Net) predicted temporal Development of a Flood Process}
%     \label{fig:temporal_unet}
% \end{figure}

% \begin{figure}
%     \centering
%     \includegraphics[width =1 \textwidth]{figures/time_series20_gan.pdf}
%     \caption{The GAN predicted temporal Development of a Flood Process}
%     \label{fig:temporal_gan}
% \end{figure}

% \begin{figure}
%     \centering
%     \includegraphics[width =1 \textwidth]{figures/time_series20_posgan.pdf}
%     \caption{The GAN with Ensemble method predicted temporal Development of a Flood Process}
%     \label{fig:temporal_ganensemble}
% \end{figure}

\begin{figure}
    \centering
    \includegraphics[width =1 \textwidth]{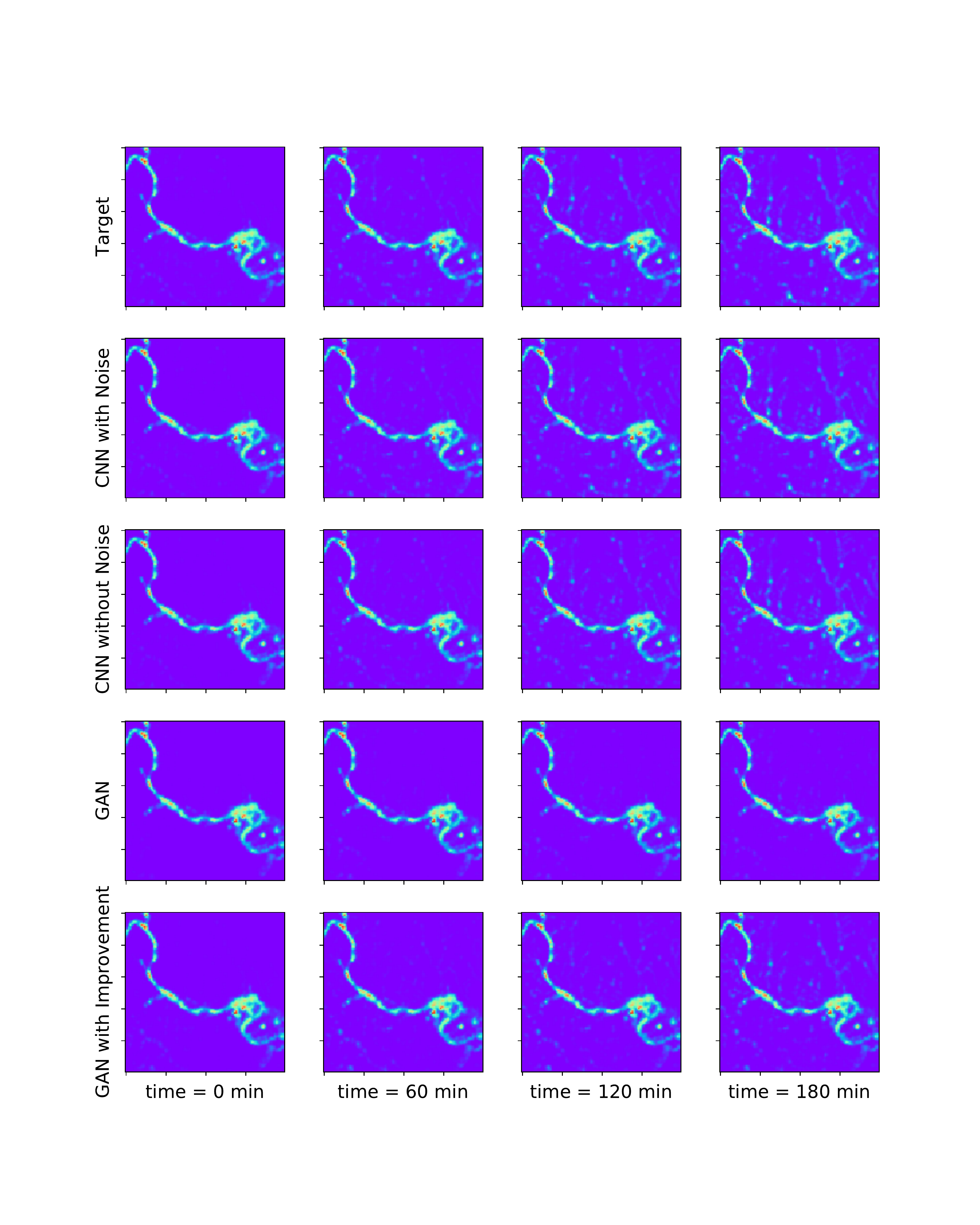}
    \caption{The comparison between targets and different deep learning based prediction methods. We compare the prediction by different methods with targets at $0$ minutes, $60$ minutes, $120$ minutes and $180$ minutes. All the methods are provided with same initial conditions and inputs. The first row is the target.}
    \label{fig:timeseries20_demo}
\end{figure}

\begin{figure}
    \centering
    \includegraphics[width = 0.8 \textwidth]{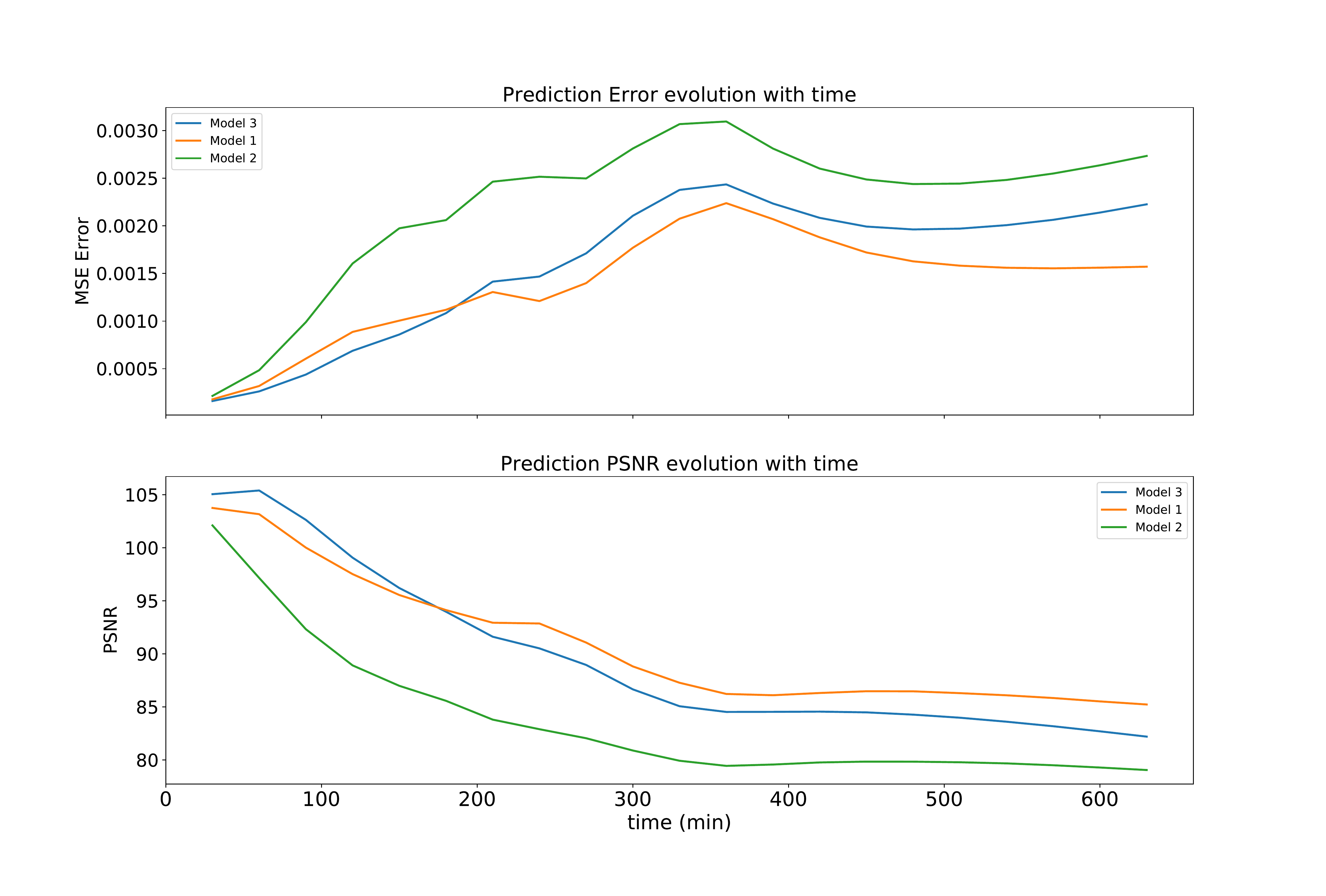}
    \caption{The comparison of performance of Model 1,2 and 3. In the first plot, the development of MSE error along time is shown. In the second plot, we present the PSNR index along time by. The results are computed under $30$ different pieces of time series data and then averaged.}
    \label{fig:CNN_general}
\end{figure}

\begin{figure}
    \centering
    \includegraphics[width = 0.8 \textwidth]{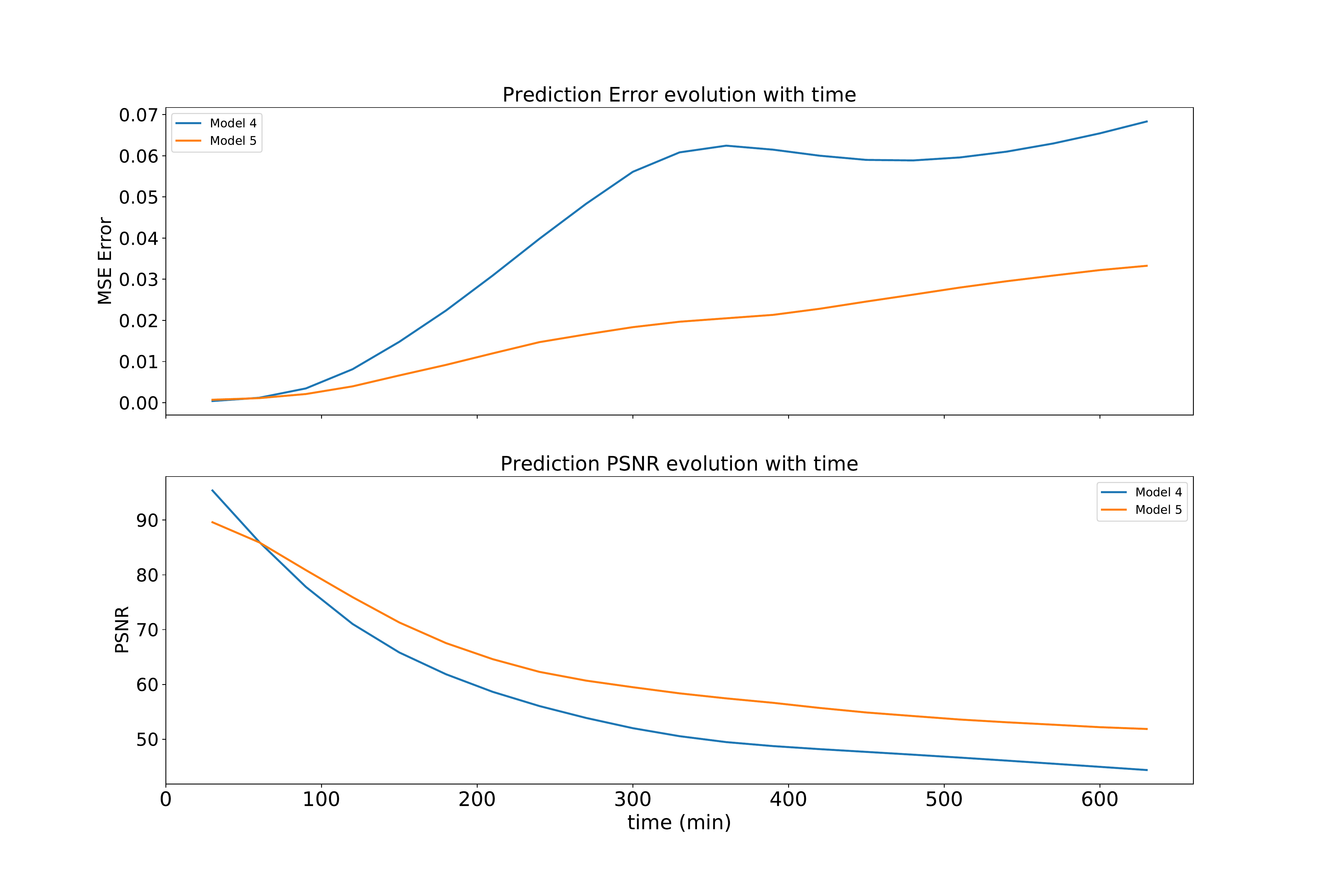}
    \caption{The comparison of performance of Model 4 and 5. In the first plot, the development of MSE error along time is shown. In the second plot, we present the PSNR index along time by. The results are computed under $30$ different pieces of time series data and then averaged.}
    \label{fig:GAN_general}
\end{figure}

First, let us discuss the results shown in Fig.~\ref{fig:timeseries20_demo}. We pick a case and predict respectively with Models 1, 2, 4 and 5. The results are plotted at time=0 minutes, 60 minutes, 120 minutes and 180 minutes. We can see that the prediction results by Model 1 and Model 2 look almost the same as the target. Meanwhile, we can see that the predictions by Model 4 totally diverge from targets in creek areas. The reason behind this has been mentioned in Section \ref{sec:methodology}. Trivial omits by conditional GANs accumulated in temporal evolution can lead to a noticeable divergence. With the improvements made by the posterior-adjustment step, Model 5 can also learn the development of flood, just not as effectively as Models 1 and 2. More tests can be found in the Appendix.

Let us then analyze the results quantitatively. In Fig.~\ref{fig:CNN_general}, we plot the averaged MSE and PSNR by Models 1, 2 and 3. Although it was shown in Table.~\ref{table:precision_general} that Model 2 leads to higher precision than Model 1 in prediction one step, it is shown in Fig.~\ref{fig:CNN_general} that Model 1 performs better in temporal evolution. This makes sense, as the noise added to training data mimics the imprecision of data-driven model caused by the previous prediction steps. Furthermore, although Model 3 performs worse than Model 1 and 2 in Table \ref{table:precision_general}, it has a similar performance as Model 1 in temporal evolution in Fig.\ref{fig:CNN_general}. This means that lowering structural loss with a discriminator can improve the performance of a data-driven model in temporal evolution. In Fig.~\ref{fig:GAN_general}, the improvements over conditional GANs by a post-adjustment step are quantitatively illustrated along the temporal evolution.

\subsection{Validation of Measurement update}
It is mentioned in Section \ref{sec:intro} that the data-driven model would be used for the real-time flood estimation purpose. It is validated in this section that the data-driven model can be applied to flood state estimation.

% In this section, we would like to check two things: would the spatial correlation still work with data-driven model? would the data-driven model robust to measurement update?

\begin{figure}
    \centering
    \includegraphics[width = 0.8 \textwidth]{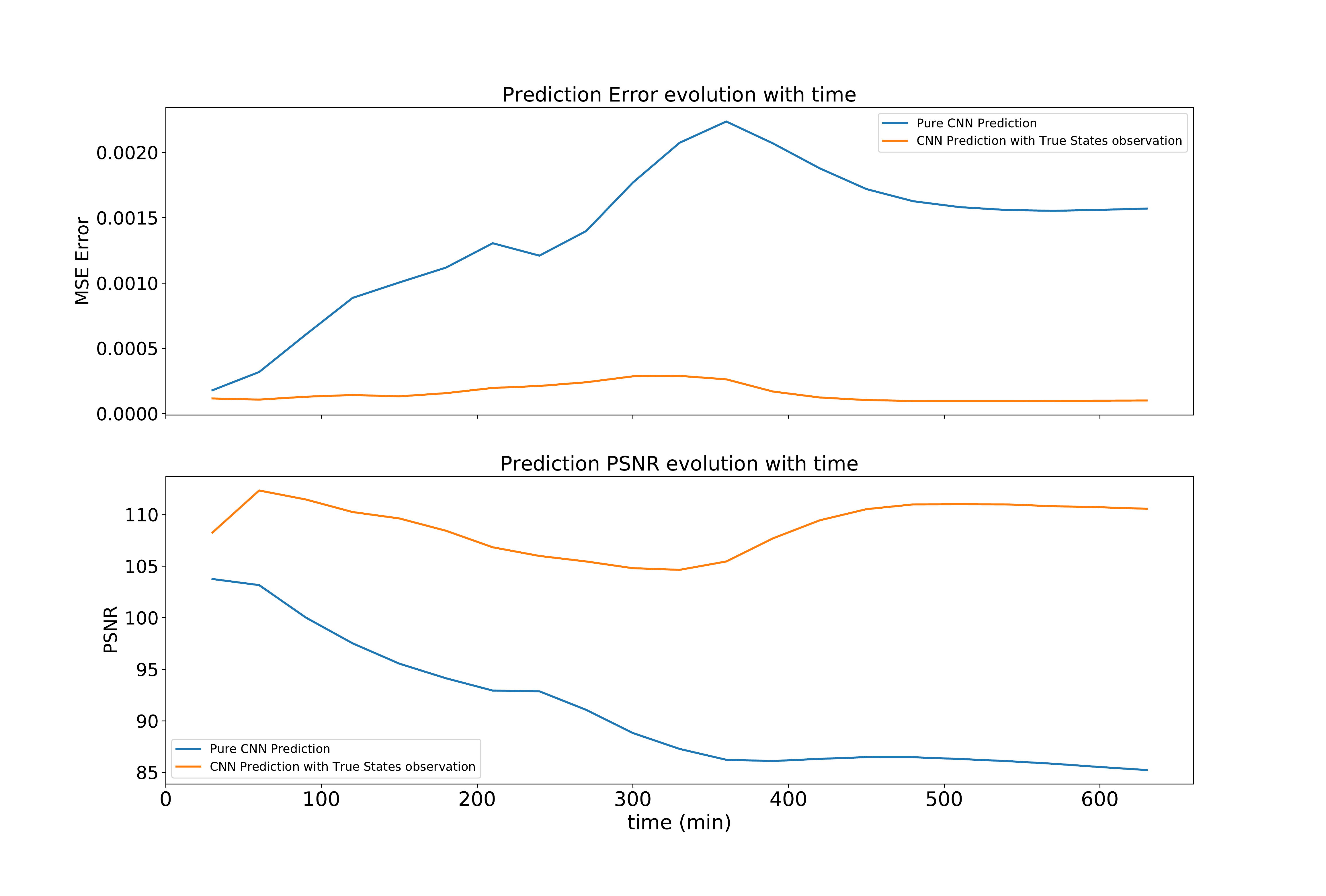}
    \caption{The comparison of performance of Model 1 and Model 1 with Measurement update from true states. In the first plot, the development of MSE error along time is shown. In the second plot, we present the PSNR index along time by. The results are computed under $30$ different pieces of time series data and then averaged.}
    \label{fig:measurement_cnn_general}
\end{figure}

Model 1 is chosen as the base model. Measurement update is conducted after each prediction step. An empirical co-variance matrix is extracted from simulation data to describe the spatial correlation. In Fig.~\ref{fig:measurement_cnn_general}, it is shown that the error is greatly reduced and the PSNR is improved with the measurement update over Model 1. It means that the data-driven model can work well for flood state estimation tasks.

% Model 1 is chosen as the base model. A We follow a Kalman Filter style to conduct the measurement update. Here, we apply an empirical covariance matrix to include the spatial correlation. This empirical covariance matrix comes from the large amount of simulation data. In order to make measurement work, we apply a Gaussian Process to empirical covariance matrix according to spatial relation between states.

% \begin{figure}
%     \centering
%     \includegraphics[width = \textwidth]{figures/error_analysis_ensemble.pdf}
%     \caption{The plot of error distributions before and after the measurement update with true states. The left shows the error distribution of the prediction model without any information of target states. The right shows the distribution of error with measurement update and information from targets.}
%     \label{fig:measurement_cnn_general}
% \end{figure}

% \begin{figure}
%     \centering
%     \includegraphics[width = \textwidth]{figures/Measurement_compare_channel.pdf}
%     \caption{The comparison of Pure CNN Prediction and CNN prediction with measurement update from targets}
%     \label{fig:measurement_cnn_channel}
% \end{figure}
\section{Conclusion}
\label{sec:conclusion}
% One characteristic of PDE data is that the scale of data varies a lot between different grids in the same channel. Thus, the forward neural network and a simple loss function cannot learn this kind of task very well. Firstly, the loss function will tend to put more weights on those data with large scale and thus neglect those data of small scale. Secondly, because simple forward network doesn't have a discriminator to judge if the prediction result is close to true result during the training process, thus the error will definitely accumulate thus leading to the instability. GAN overcome these two difficulties. Firstly, the discriminator can be viewed as a loss function which is dynamically learned that can treat data with different scales equally. Secondly, the discriminator will bring the prediction to the distribution thus preventing the instability from happening.
In this work, we develop physics-informed, data-driven models to predict the evolution of a two dimensional flood in a test area of Austin, Texas. The data-driven model recovers the dynamics encoded by the 2D SWE equations and boosts the computation speed by several orders of magnitude (approximately $\times50,000$). Compared to statistics-based and other simplified flood prediction models, the physics-informed, data-driven model provides detailed and real-time city-wide flood development predictions. This can serve as the prediction model for real-time flood monitoring, estimation, or model-predictive control applications.

Several Deep Learning techniques, including CNN and conditional GANs, are applied in constructing the learning structure and the loss functions. According to this study, appropriate CNN architectures can recover the dynamics of the 2D SWE very well. Conditional GANs can reduce the structural error of the prediction with the discriminator, though it can cause a divergence over longer time horizons. In order to overcome this divergence, an assimilation step inspired from the update step of the Kalman Filter is introduced and proves to be useful. Adding more weights to the $L_1$ norm term in the loss function of conditional GANs can also improve the performance. Different models are compared both qualitatively and quantitatively in Section \ref{sec:result}. It is shown that the predictions made by a data-driven model aligns well with the results given by the SWE simulation.

% The satisfying precision means that the data-driven model can carry the dynamics governed by SWE and be used for 2-D real-time flood prediction work.

Future works will involve the following aspects. First, state estimation techniques based on the physics-informed, data-driven model for flood estimation can be developed (i.e. using a classical Ensemble Kalman Filter or a Particle Filter). Second, the influence of different parameter sets with SWE on data-driven models should be studied. The parameters of the SWE, including elevation and friction coefficients, will greatly affect the outcome of the simulation. Thus, this scheme can be used to perform inverse-modeling (parameter estimation) based on actual flood data. Third, more advanced and realistic simulations can be configured to provide training data for the data-driven model. In this work, although a complex setting of an urban area with realistic topology information is studied, more detailed settings have not been specified in the model. For example, the saturation of the soil by the flood (which affects the model parameters), the sewers in urban areas, and the local impact of the dense buildings in the downtown areas have not been precisely modeled, other than by their low runoff coefficients.

\section*{References}
\bibliographystyle{elsarticle-harv}
\bibliography{sample}

\newpage
\section*{Appendix}
\label{sec:appendix}
\subsection*{Training Details}

\subsubsection*{CNN}

\begin{figure}
    \centering
    \includegraphics[width=\textwidth]{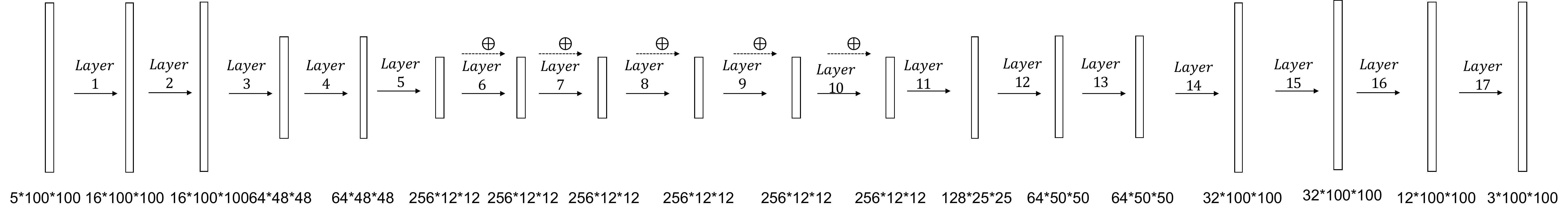}
    \caption{The general Architecture of Residual Block based Network. For the details in each layer, please refer to corresponding table in Appendix. Table\ref{table:ResDetails}}
    \label{fig:resStruct}
\end{figure}

Following the basic idea of Section \ref{sec:methodology}, we tested several architectures and chose the highest performing. The architecture finally adopted is shown in Fig.\ref{fig:resStruct}. The $100\times100$ dimension is firstly scaled down and then some residual blocks are applied. The dimension is then scaled up back to $100\times100$. The details(activation function, kernel size in each layer) can be found in Table \ref{table:ResDetails}.
$8000$ epochs are run for each training task. We tested the performance of Stochastic Gradient Descent (SGD) and ADAM \cite{kingma2014adam}. It is found that SGD and ADAM will eventually have similar testing and training errors, though ADAM has a much faster convergence rate at the beginning epochs. In terms of learning rate schedule, several popular schedules are tested, including fixed learning rate, decaying with epoch($\frac{0.001}{\sqrt{epoch num}}$) and periodic learning rate.

% After normalizing the data, the performances of inputs with or without Gaussian noise added are compared. Since enough amount of training data has been supplied by simulations, methods used to avoid overfitting such as dropout do not improve the results much.

% \begin{figure}
%     \centering
%     \includegraphics[width=\textwidth]{figures/UNetStruct.pdf}
%     \caption{The general Architecture of UNet. For the details in each layer, please refer to corresponding table in Appendix. Table\ref{table:UNetDetails}}
%     \label{fig:UNetStruct}
% \end{figure}

\begin{table}
\centering
\begin{tabular}{|l|l|} 
\hline
layer 1  &      \shortstack{ channel: $5 \rightarrow 16$; kernel size: $5\times 5$; Stride: $1$; Padding: $0$;\\ Batch Normalization; Activation function: Parametric ReLUs}    \\ 
\hline
layer 2  &       \shortstack{channel: $16 \rightarrow 16$; kernel size: $1\times 1$; No padding;\\ Batch Normalization; Activation function: Parametric ReLUs}     \\ 
\hline
layer 3  &       \shortstack{channel: $16 \rightarrow 64$; kernel size: $2\times 2$; Stride: $2$; Padding: $0$;\\ Batch Normalization; Activation function: Parametric ReLUs}     \\ 
\hline
layer 4  &       \shortstack{channel: $64 \rightarrow 64$; kernel size: $1\times 1$; Stride: $1$; Padding: 0;\\ Batch Normalization; Activation function: Parametric ReLUs}     \\ 
\hline
layer 5  &       \shortstack{channel: $64 \rightarrow 256$; kernel size: $4\times 4$; stride : $4$; No padding;\\ Batch Normalization; Activation function: Parametric ReLUs}     \\ 
\hline
layer 6  &       \shortstack{channel: $256 \rightarrow 256$; kernel size: $1\times 1$; No padding;\\ Batch Normalization; Activation function: Parametric ReLUs}     \\ 
\hline
layer 7  &       \shortstack{channel: $256 \rightarrow 256$; kernel size: $3\times 3$; Padding: 1;\\ Batch Normalization; Activation function: Parametric ReLUs}     \\ 
\hline

layer 8  &       \shortstack{channel: $256 \rightarrow 256$; kernel size: $3\times 3$; Padding: 1; \\ Batch Normalization; Activation function: Parametric ReLUs}     \\ 
\hline

layer 9  &       \shortstack{channel: $256 \rightarrow 256$; kernel size: $3\times 3$; Padding: 1;\\ Batch Normalization; Activation function: Parametric ReLUs}     \\ 
\hline

layer 10  &       \shortstack{channel: $256 \rightarrow 256$; kernel size: $3\times 3$; Padding: 1;\\ Batch Normalization; Activation function: Parametric ReLUs}     \\ 
\hline

layer 11  &       \shortstack{channel: $256 \rightarrow 128$; kernel size: $3\times 3$; Stride: 2;\\ Batch Normalization; Activation function: Parametric ReLUs}     \\ 
\hline

layer 12  &       \shortstack{channel: $128 \rightarrow 64$; kernel size: $2\times 2$; Stride: 2;\\ Batch Normalization; Activation function: Parametric ReLUs}     \\ 
\hline

layer 13  &       \shortstack{channel: $64 \rightarrow 64$; kernel size: $1\times 1$; Padding: 0;\\ Batch Normalization; Activation function: Parametric ReLUs}     \\ 
\hline

layer 14  &       \shortstack{channel: $64 \rightarrow 32$; kernel size: $2\times 2$; Stride: 2;\\ Batch Normalization; Activation function: Parametric ReLUs}     \\ 
\hline

layer 15  &       \shortstack{channel: $32 \rightarrow 32$; kernel size: $1\times 1$; Stride: 1; Padding: 0;\\ Batch Normalization; Activation function: Parametric ReLUs}     \\ 
\hline

layer 16  &       \shortstack{channel: $32 \rightarrow 12$; kernel size: $3\times 3$; Stride: 1; Padding: 1;\\ Batch Normalization; Activation function: Parametric ReLUs}     \\ 
\hline

layer 17  &       \shortstack{channel: $12 \rightarrow 3$; kernel size: $1\times 1$; Stride: 1; Padding: 0;\\ Batch Normalization; Activation function: Parametric ReLUs}     \\ 
\hline
\end{tabular}
\caption{The Details for ResBlock based CNN}
\label{table:ResDetails}
\end{table}

\subsubsection*{GANs}

For GANs models, we follow the same architecture as the one used for CNNs in previous section for the generator and Patch CNN~\cite{mirza2014conditional} for the discriminator. We sued RMSprop as optimized, trained for 10000 epochs on the dataset. Both generator and discriminator had learning rate of 0.0002. The fake and true labels were soft labels in range(0.9,1.0) and range(0.0,0.1) respectively.

\subsection*{More Illustrations}

\begin{figure}
    \centering
    \includegraphics[width =1 \textwidth]{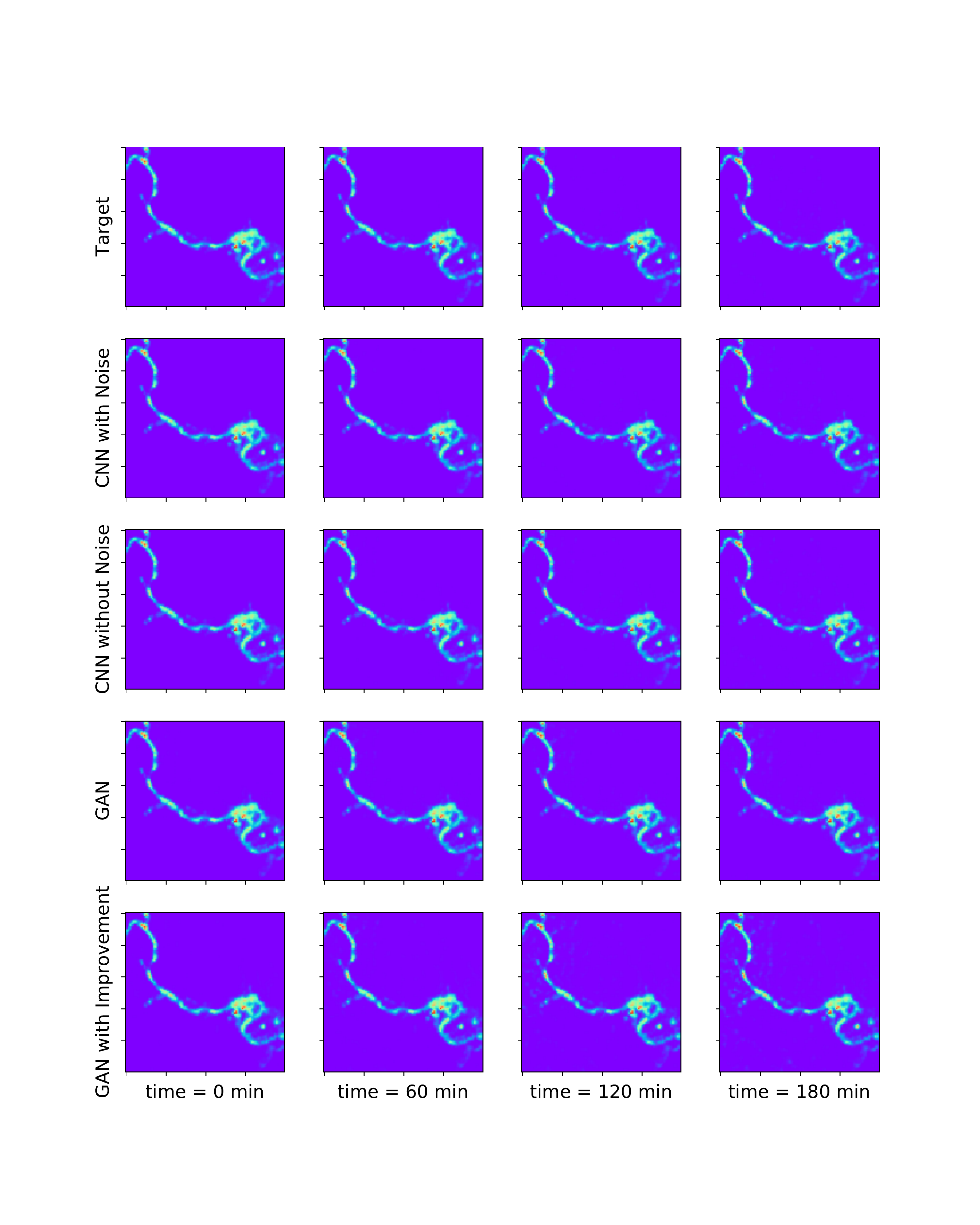}
    \caption{The comparison between targets and different deep learning based prediction methods. We compare the prediction by different methods with targets at $0$ minutes, $60$ minutes, $120$ minutes and $180$ minutes. All the methods are provided with same initial conditions and inputs. The first row is the target.}
    \label{fig:timeseries_demo_appendix_1}
\end{figure}

\begin{figure}
    \centering
    \includegraphics[width =1 \textwidth]{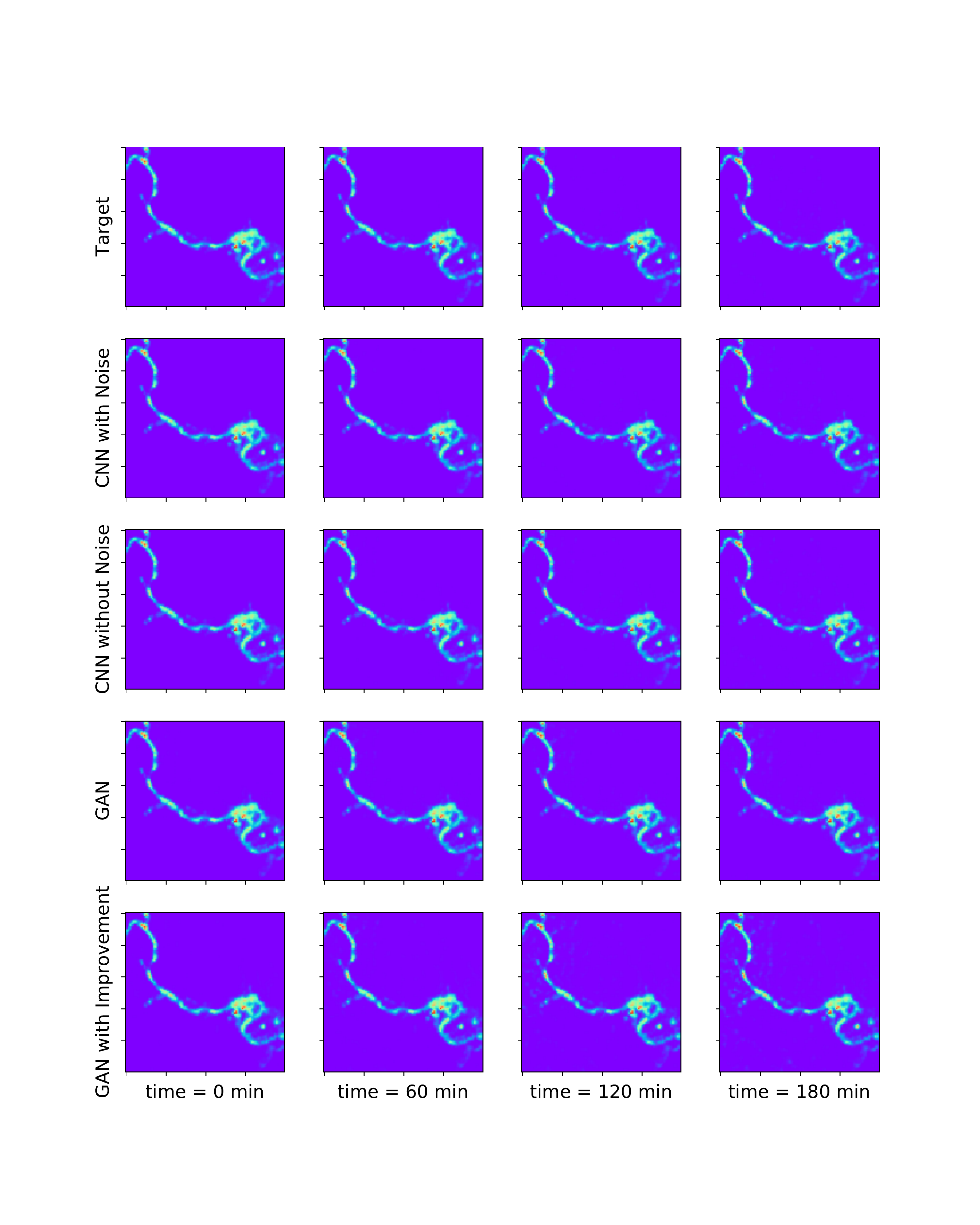}
    \caption{The comparison between targets and different deep learning based prediction methods. We compare the prediction by different methods with targets at $0$ minutes, $60$ minutes, $120$ minutes and $180$ minutes. All the methods are provided with same initial conditions and inputs. The first row is the target.}
    \label{fig:timeseries_demo_appendix_1}
\end{figure}

\begin{figure}
    \centering
    \includegraphics[width =1 \textwidth]{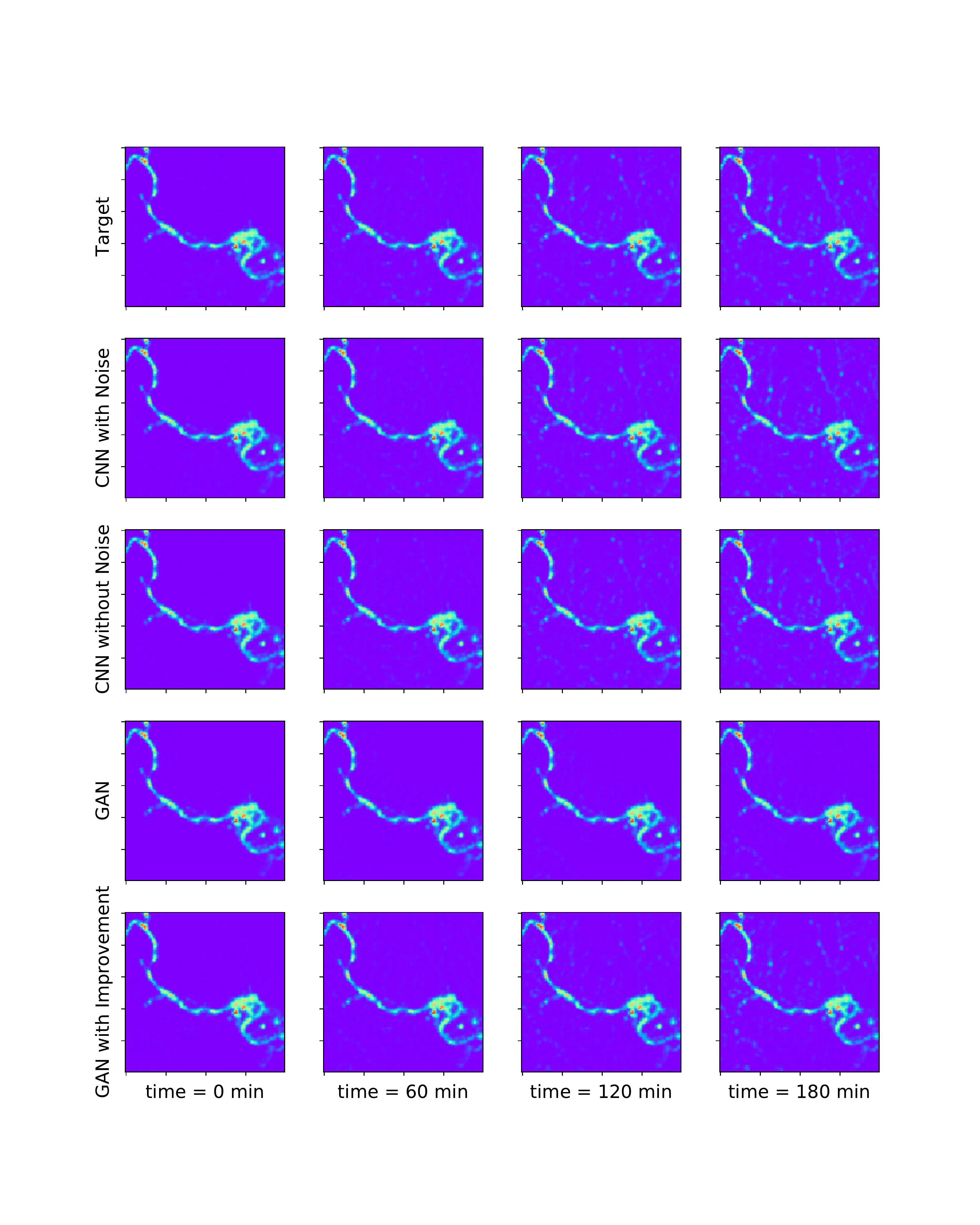}
    \caption{The comparison between targets and different deep learning based prediction methods. We compare the prediction by different methods with targets at $0$ minutes, $60$ minutes, $120$ minutes and $180$ minutes. All the methods are provided with same initial conditions and inputs. The first row is the target.}
    \label{fig:timeseries_demo_appendix_2}
\end{figure}

\begin{figure}
    \centering
    \includegraphics[width =1 \textwidth]{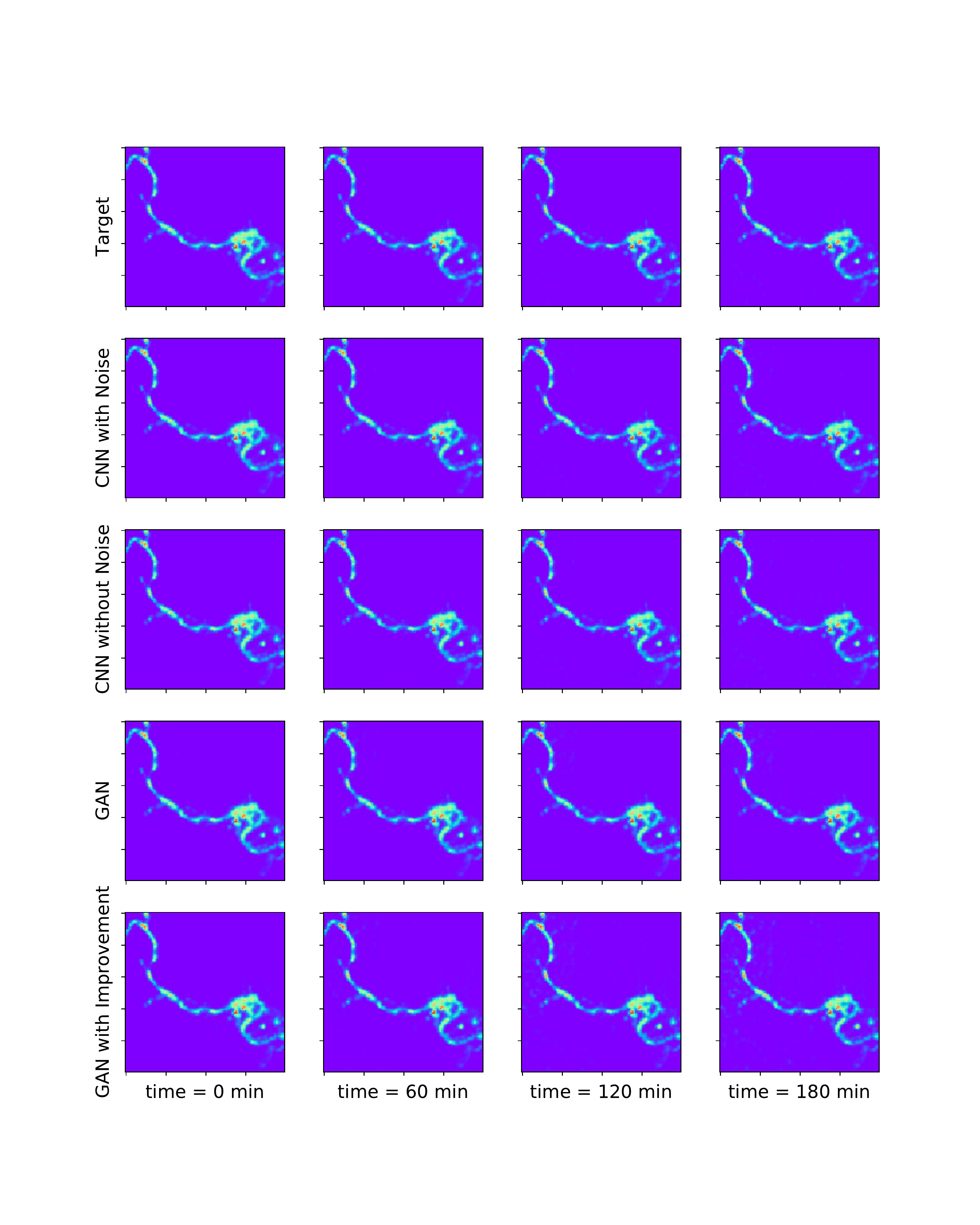}
    \caption{The comparison between targets and different deep learning based prediction methods. We compare the prediction by different methods with targets at $0$ minutes, $60$ minutes, $120$ minutes and $180$ minutes. All the methods are provided with same initial conditions and inputs. The first row is the target.}
    \label{fig:timeseries_demo_appendix_3}
\end{figure}

\begin{figure}
    \centering
    \includegraphics[width =1 \textwidth]{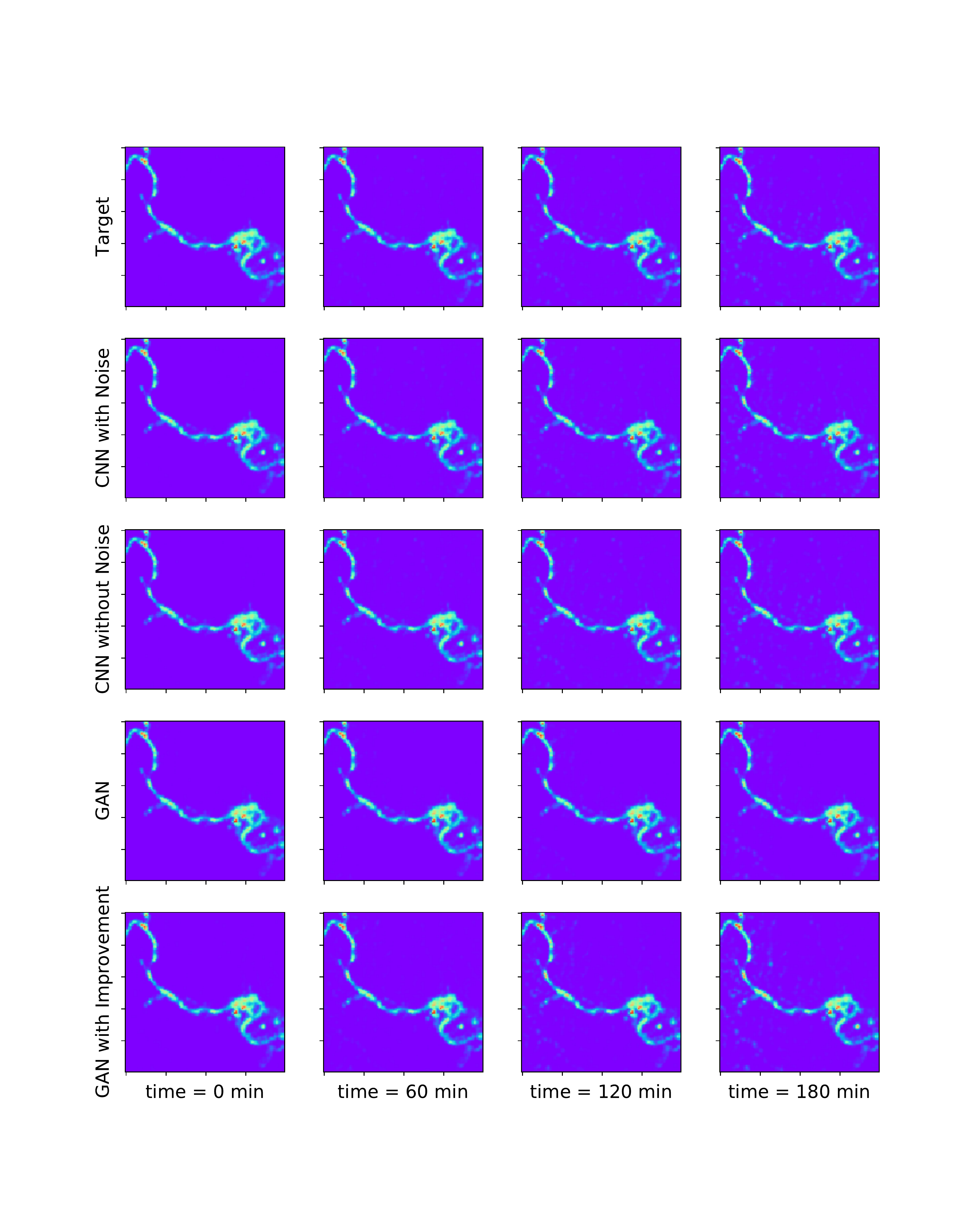}
    \caption{The comparison between targets and different deep learning based prediction methods. We compare the prediction by different methods with targets at $0$ minutes, $60$ minutes, $120$ minutes and $180$ minutes. All the methods are provided with same initial conditions and inputs. The first row is the target.}
    \label{fig:timeseries_demo_appendix_4}
\end{figure}

\begin{figure}
    \centering
    \includegraphics[width =1 \textwidth]{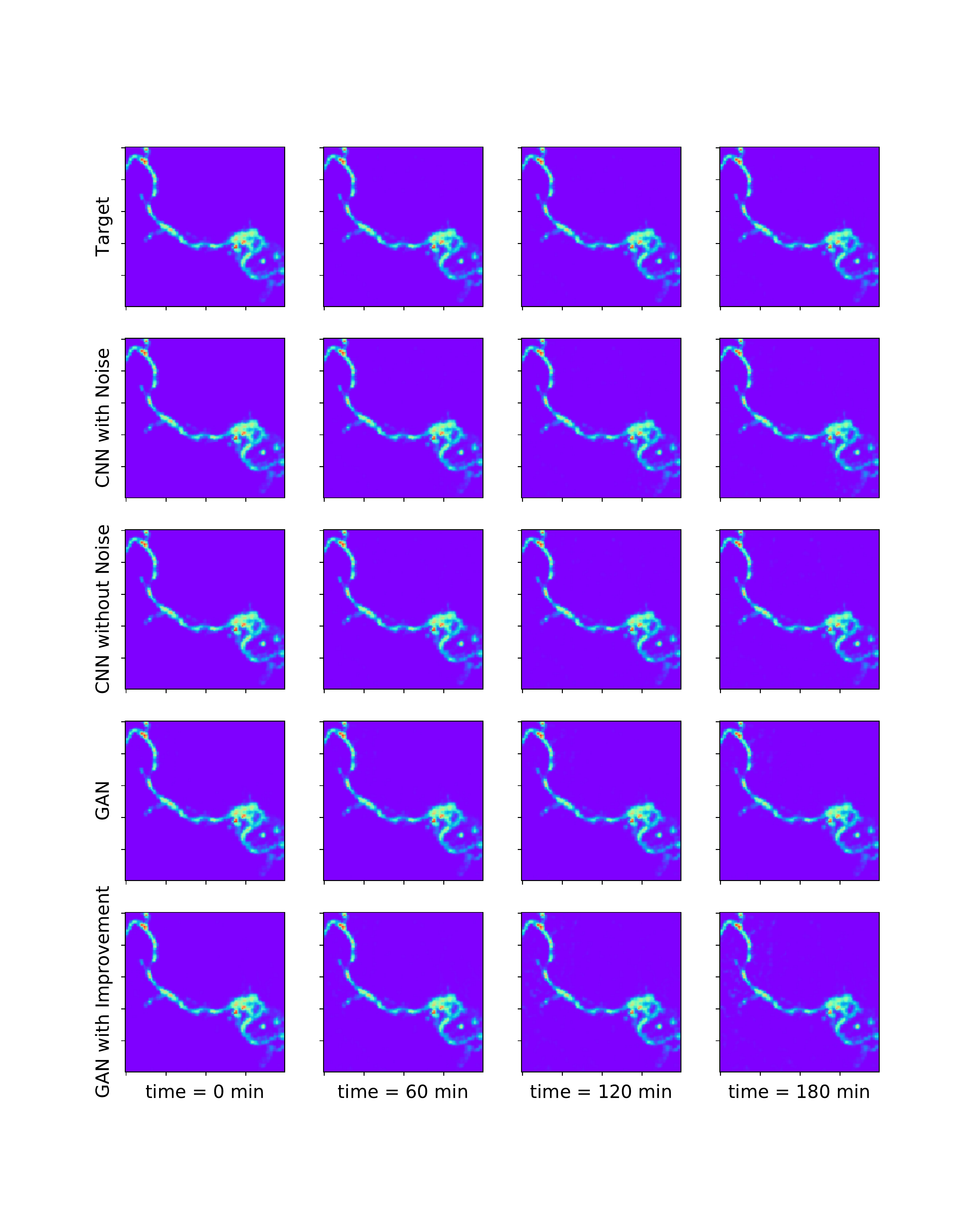}
    \caption{The comparison between targets and different deep learning based prediction methods. We compare the prediction by different methods with targets at $0$ minutes, $60$ minutes, $120$ minutes and $180$ minutes. All the methods are provided with same initial conditions and inputs. The first row is the target.}
    \label{fig:timeseries_demo_appendix_5}
\end{figure}

\begin{figure}
    \centering
    \includegraphics[width =1 \textwidth]{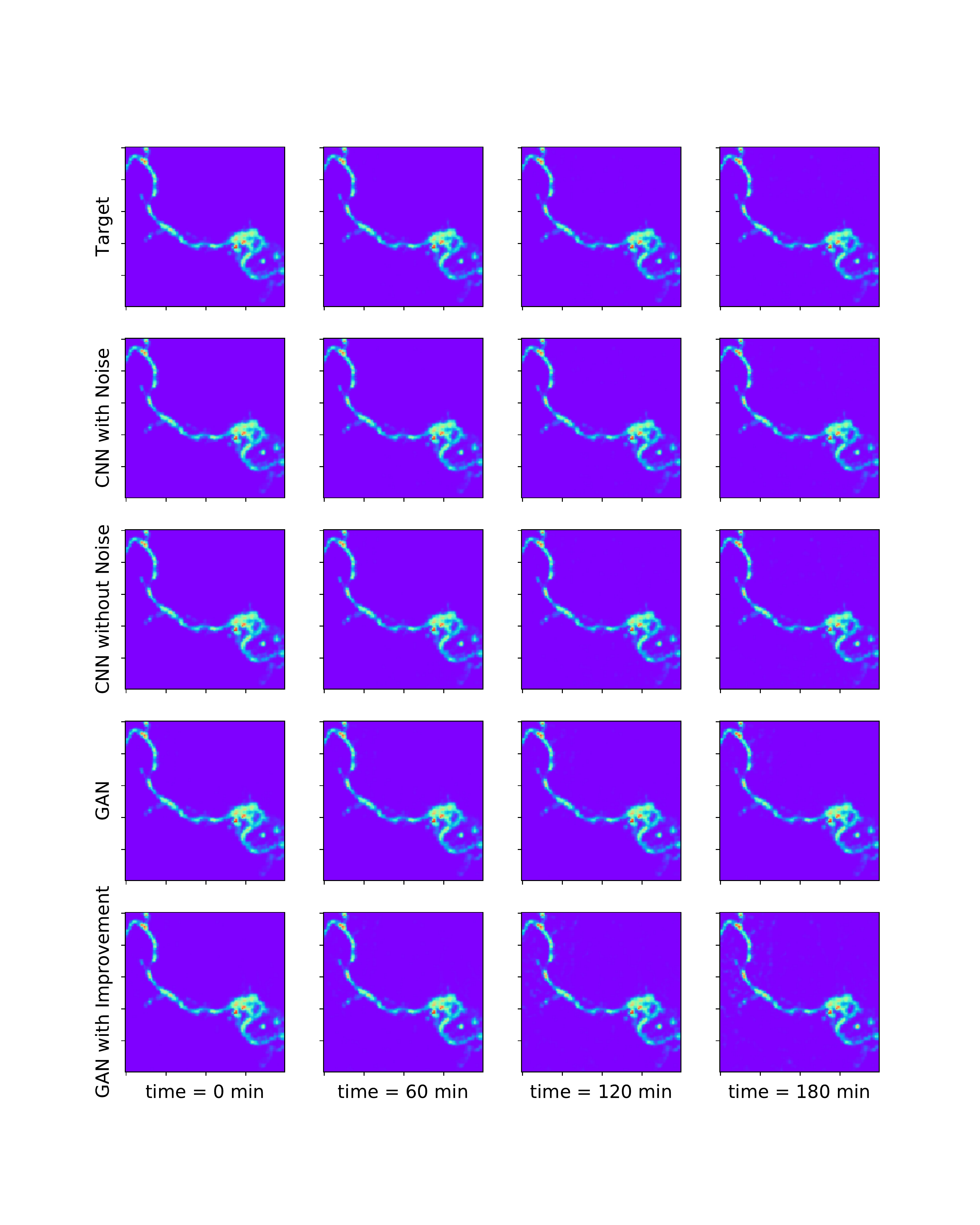}
    \caption{The comparison between targets and different deep learning based prediction methods. We compare the prediction by different methods with targets at $0$ minutes, $60$ minutes, $120$ minutes and $180$ minutes. All the methods are provided with same initial conditions and inputs. The first row is the target.}
    \label{fig:timeseries_demo_appendix_6}
\end{figure}

\begin{figure}
    \centering
    \includegraphics[width =1 \textwidth]{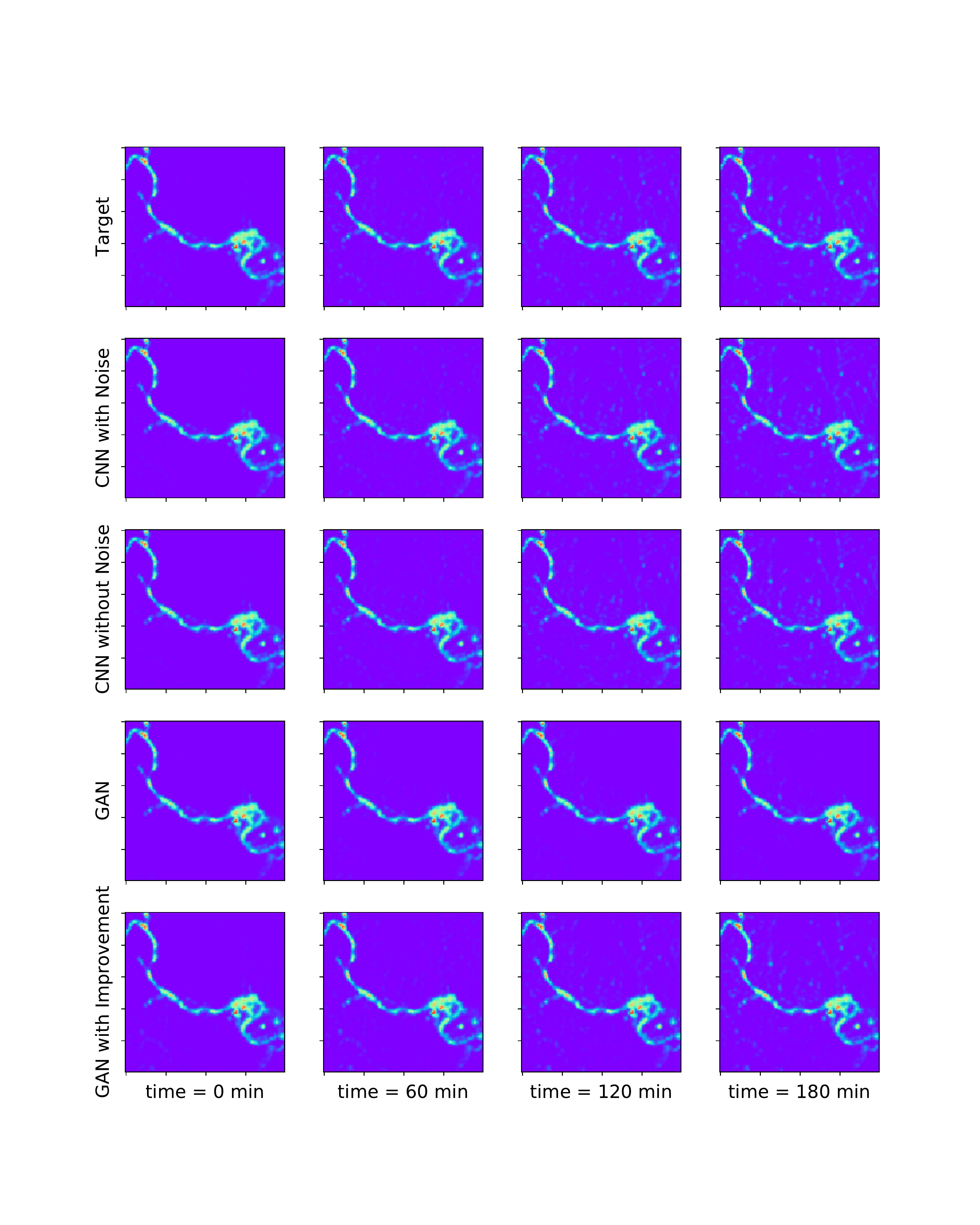}
    \caption{The comparison between targets and different deep learning based prediction methods. We compare the prediction by different methods with targets at $0$ minutes, $60$ minutes, $120$ minutes and $180$ minutes. All the methods are provided with same initial conditions and inputs. The first row is the target.}
    \label{fig:timeseries_demo_appendix_7}
\end{figure}

\begin{figure}
    \centering
    \includegraphics[width =1 \textwidth]{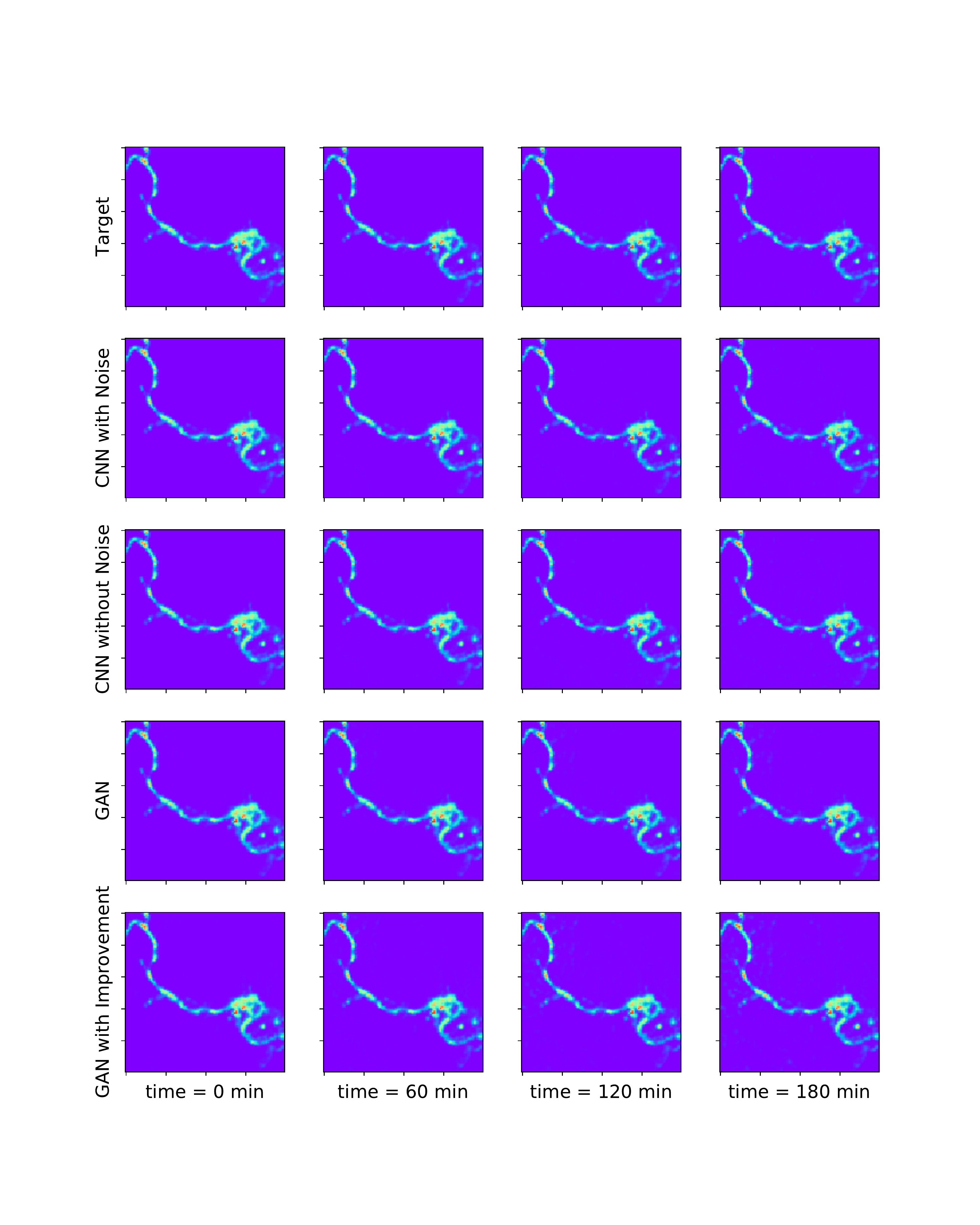}
    \caption{The comparison between targets and different deep learning based prediction methods. We compare the prediction by different methods with targets at $0$ minutes, $60$ minutes, $120$ minutes and $180$ minutes. All the methods are provided with same initial conditions and inputs. The first row is the target.}
    \label{fig:timeseries_demo_appendix_8}
\end{figure}

\begin{figure}
    \centering
    \includegraphics[width =1 \textwidth]{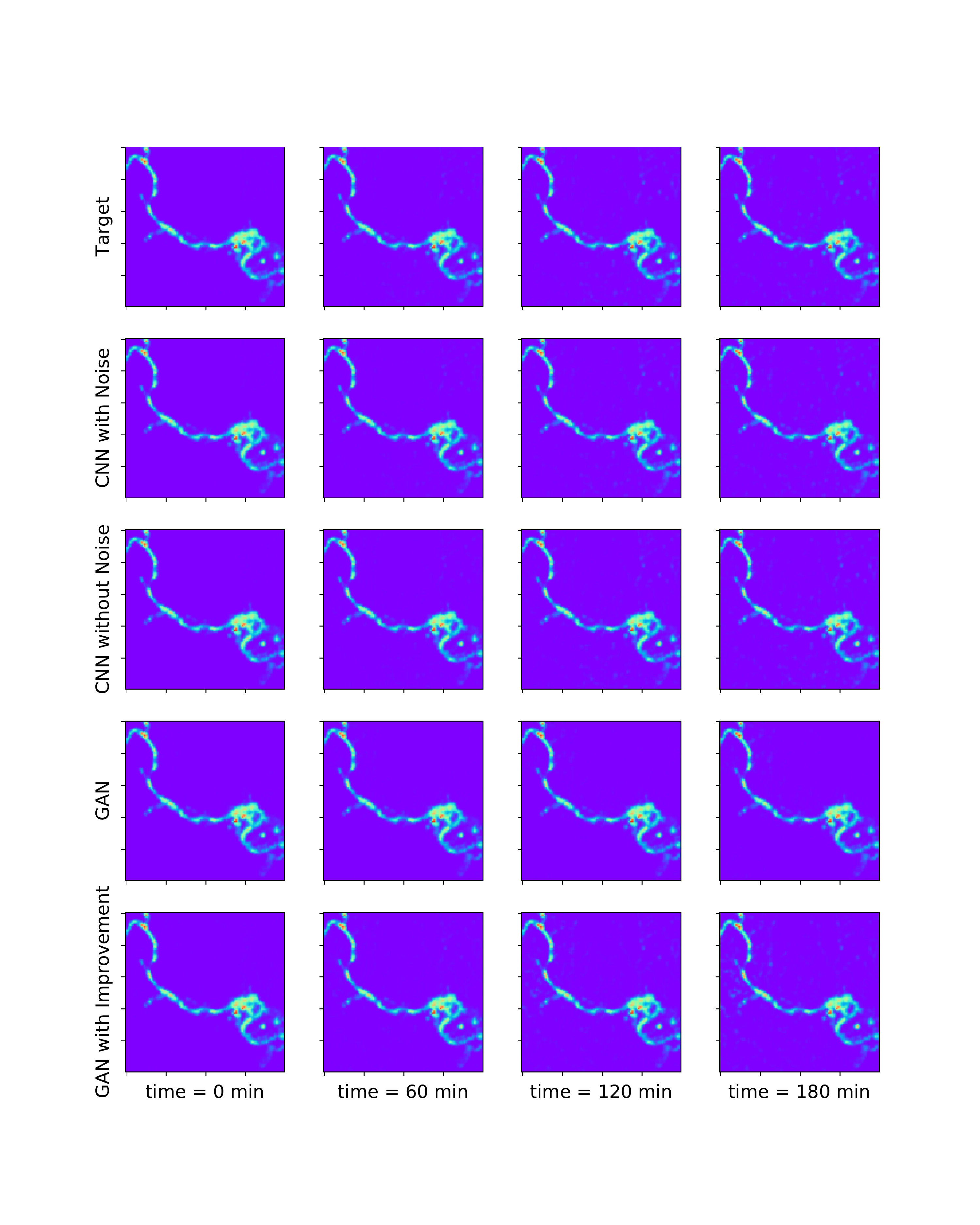}
    \caption{The comparison between targets and different deep learning based prediction methods. We compare the prediction by different methods with targets at $0$ minutes, $60$ minutes, $120$ minutes and $180$ minutes. All the methods are provided with same initial conditions and inputs. The first row is the target.}
    \label{fig:timeseries_demo_appendix_9}
\end{figure}

\begin{figure}
    \centering
    \includegraphics[width =1 \textwidth]{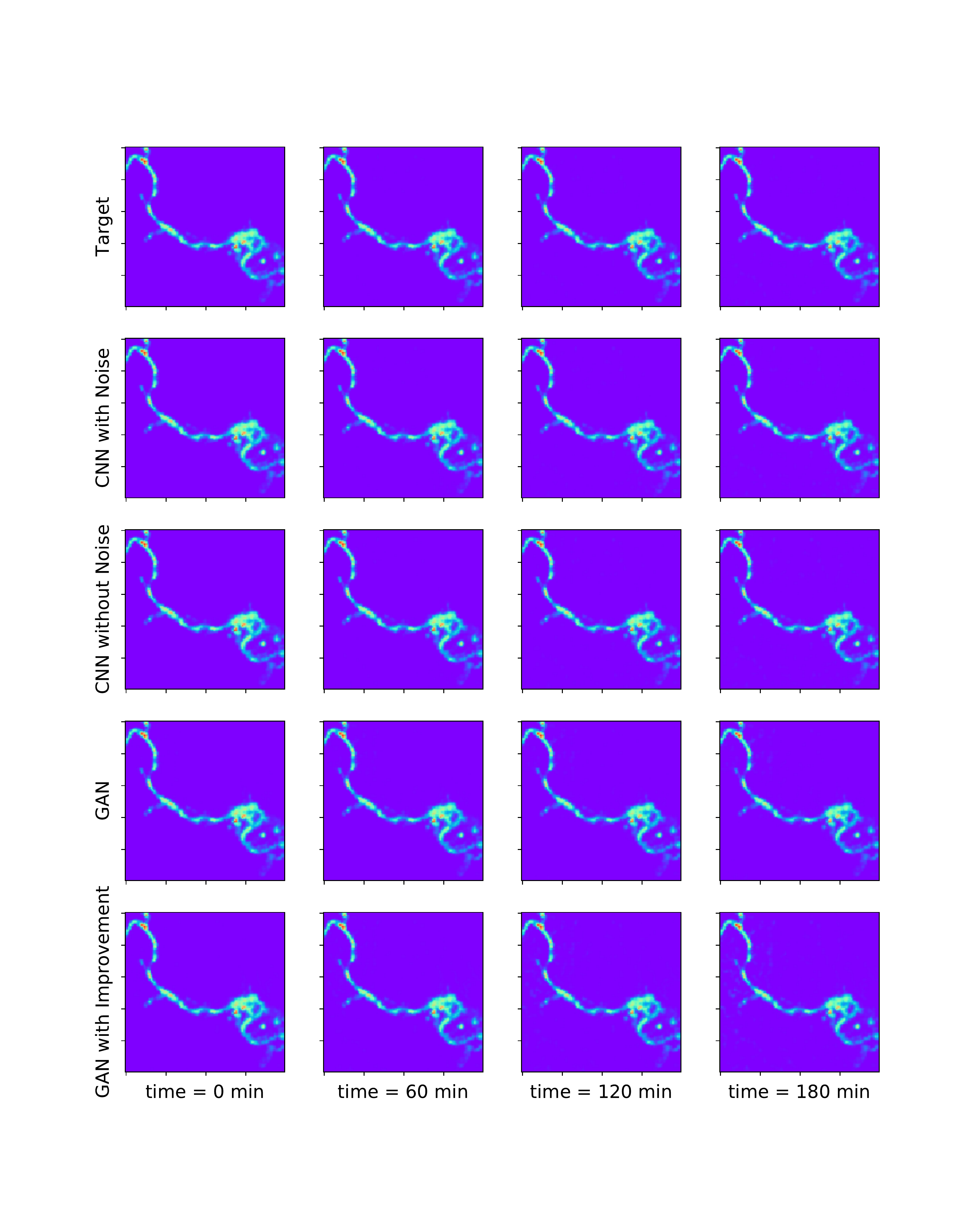}
    \caption{The comparison between targets and different deep learning based prediction methods. We compare the prediction by different methods with targets at $0$ minutes, $60$ minutes, $120$ minutes and $180$ minutes. All the methods are provided with same initial conditions and inputs. The first row is the target.}
    \label{fig:timeseries_demo_appendix_10}
\end{figure}
\end{document}